\documentclass[aps,prl,twocolumn]{revtex4-1}

\usepackage{geometry}
\usepackage{setspace}
\usepackage{bm}
\usepackage{mathtools}
\usepackage{amssymb}
\usepackage{float}
\usepackage{xr}
\usepackage{xcolor}

\begin{document}

\title{Flat-band localization and interaction-induced delocalization of photons}
\author{Jeronimo~G.C.~Martinez$^*$}
\affiliation{Princeton University, Department of Electrical and Computer Engineering}
\author{Christie~S.~Chiu$^*$}
\affiliation{Princeton University, Department of Electrical and Computer Engineering}
\author{Basil~M.~Smitham}
\affiliation{Princeton University, Department of Electrical and Computer Engineering}
\author{Andrew~A.~Houck}
\affiliation{Princeton University, Department of Electrical and Computer Engineering}

\begin{abstract}
    Advances in quantum engineering have enabled the design, measurement, and precise control of synthetic condensed matter systems \cite{altman_quantum_2021}. 
    The platform of superconducting circuits offers two particular capabilities \cite{carusotto_photonic_2020, leykam_artificial_2018}: flexible connectivity of circuit elements that enables a variety of lattice geometries, and circuit nonlinearity that provides access to strongly interacting physics. 
    Separately, these features have allowed for the creation of curved-space lattices \cite{kollar_hyperbolic_2019} and the realization of strongly correlated phases \cite{roushan_spectroscopic_2017, ma_dissipatively_2019, saxberg_disorder-assisted_2022} and dynamics \cite{yan_strongly_2019, gong_quantum_2021, karamlou_quantum_2022} in one-dimensional chains and square lattices.
    Missing in this suite of simulations is the simultaneous integration of interacting particles into lattices with unique band dispersions, such as dispersionless flat bands.
    An ideal building block for flat-band physics is the Aharonov-Bohm cage \cite{vidal_aharonov-bohm_1998}: a single plaquette of a lattice whose band structure consists entirely of flat bands.
    Here, we experimentally construct an Aharonov-Bohm cage and observe the localization of a single photon, the hallmark of all-bands-flat physics.
    Upon placing an interaction-bound photon pair into the cage, we see a delocalized walk indicating an escape from Aharonov-Bohm caging \cite{vidal_interaction_2000}.
    We further find that a variation of caging persists for two particles initialized on opposite sites of the cage.
    These results mark the first experimental observation of a quantum walk that becomes delocalized due to interactions and establish superconducting circuits for studies of flat-band-lattice dynamics with strong interactions.
\end{abstract}

\maketitle

Flat electronic bands quench the kinetic energy of electrons and provide a lattice environment that is uniquely susceptible to interactions, disorder, and particle statistics.
As a result, they can host a wide range of phenomena, from itinerant ferromagnetism in spinful systems \cite{lieb_two_1989, tasaki_nagaokas_1998} to the fractional quantum Hall effect \cite{von_klitzing_quantum_2017}, fractional Chern insulator states \cite{bergholtz_topological_2013, parameswaran_fractional_2013, spanton_observation_2018}, and strongly correlated phases observed in magic-angle twisted bilayer graphene \cite{cao_unconventional_2018, xie_fractional_2021}.
The role of flat bands in high-temperature superconductivity \cite{peotta_superfluidity_2015, julku_quantum_2021} and quantum thermalization \cite{goda_inverse_2006, danieli_many-body_2020, kuno_multiple_2021}, as well as searches for novel flat-band materials \cite{regnault_catalogue_2022}, are also active areas of research.

\begin{figure*}[t!]
  \begin{minipage}[c]{0.5\textwidth}
    \includegraphics[width=0.92\textwidth]{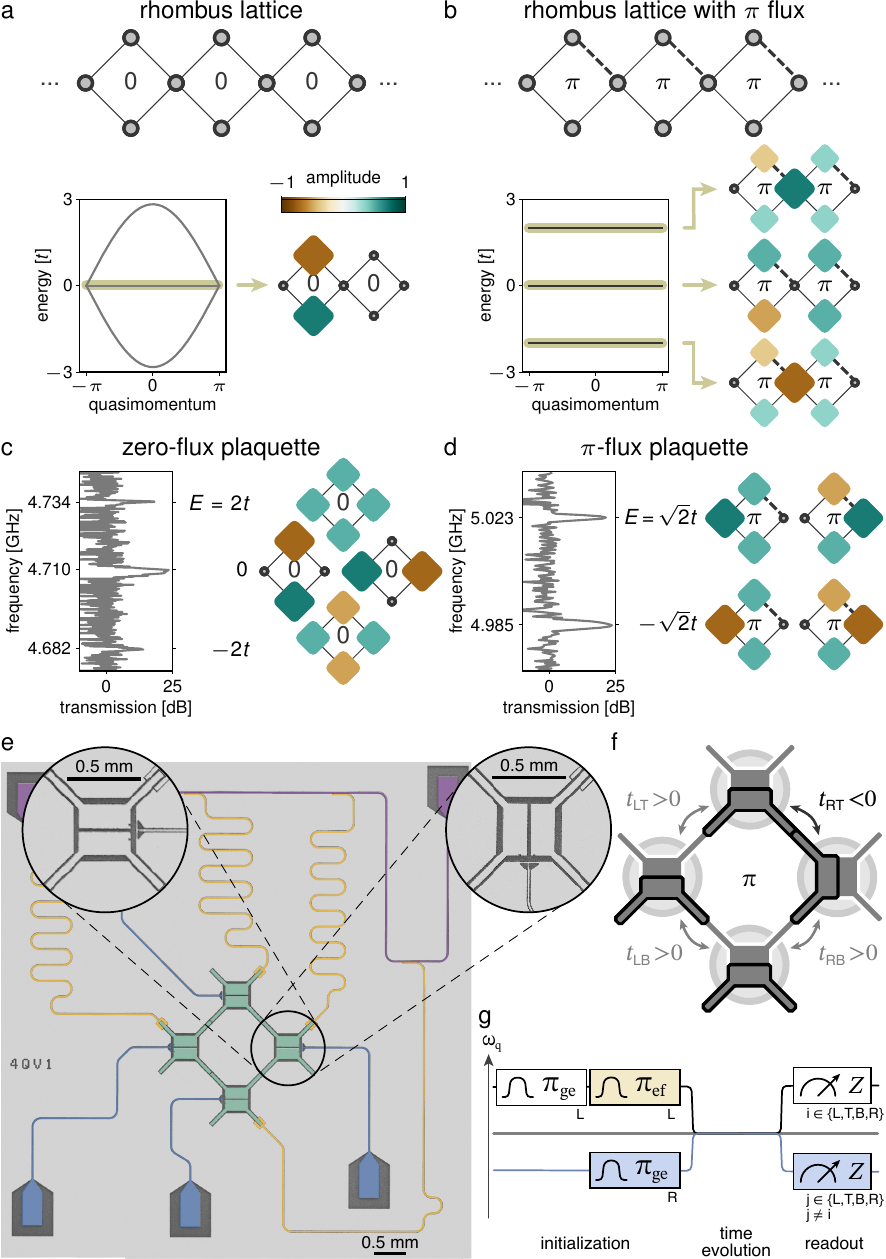}
  \end{minipage}
  \begin{minipage}[c]{0.49\textwidth}
    \caption{
       Flat-band lattices and Aharonov-Bohm caging with transmon qubits.
    \textbf{(a)} The rhombus lattice with its band structure and the depiction of a compact localized state.
    \textbf{(b)} By adding a $\pi$-flux synthetic magnetic field, all bands become flat and are spanned by compact localized states. The dashed lattice bonds indicate tunneling of opposite sign from the other tunnel couplings.
    \textbf{(c)} Eigenbasis for a single plaquette of the zero-flux rhombus lattice. 
    (left) Measured two-tone spectroscopy with all qubits on resonance. 
    (right) Eigenenergies and eigenstates calculated using exact diagonalization.
    \textbf{(d)} Same as (c) but for the  $\pi$-flux rhombus plaquette.
    \textbf{(e)} False-color image of zero-flux plaquette device, with zoomed insets highlighting the difference between the zero- (left) and $\pi$-flux (right) plaquette designs.
    \textbf{(f)} We rotate the rightmost qubit by 90 degrees to introduce a geometric phase of $\pi$. This rotation modifies how upper (white outline) and lower (black outline) electrodes are capacitively coupled and changes the sign of the upper right tunneling term.
    \textbf{(g)} Experimental sequences used in this work, which consist of initialization, time evolution with all qubits on resonance, and readout. In the first experiment (white boxes), we initialize the left qubit in the first excited state and detune one of the four qubits for readout. In the second (white and yellow boxes), we instead initialize the left qubit in the second excited state with two sequential pulses. In the third experiment (white and blue boxes), we initialize the left and right qubits in the first excited state and detune two of the four qubits for simultaneous readout. In all experiments, we vary the qubits that are detuned for readout to extract average populations for each qubit.
    In (f) and (g), subscripts L, T, B, R refer to the left, top, bottom, and right qubits, respectively.
    } 
    \label{fig:setup}
    \end{minipage}
\end{figure*}

Many types of flat bands have been theoretically explored and classified, frequently within the tight-binding model of solids where flat bands can result from specific features in lattice geometry \cite{sutherland_localization_1986}.
A simple example is the one-dimensional rhombus lattice, which contains three sites per unit cell and exhibits a flat middle band that touches dispersive bands above and below (see Fig.\,\ref{fig:setup}a).
As with many flat bands, the eigenstates within the band can be written in a basis of real-space eigenfunctions known as compact localized states \cite{aoki_hofstadter_1996}.
These states are characterized by a strict boundedness on the number of sites involved in the particle wavefunction, reflecting stronger localization than the more typical exponential localization of Wannier functions.
At the same time, because dispersive bands are present in the band structure, an initially localized particle generically contains contributions from dispersive states and will delocalize across the lattice over time.

A type of extreme localization, in which \emph{every} band of the spectrum is flat, emerges in select lattices and specific values of magnetic flux, including the rhombus lattice with $\pi$ flux.
The effect of magnetic flux on localization depends heavily on lattice geometry and can be understood through the Aharonov-Bohm effect.
More specifically, the tight-binding electron accumulates an Aharonov-Bohm phase as it hops around a closed loop of the lattice, or plaquette.
Additionally, the lattice plaquettes are arranged in such a way that causes complete destructive wavefunction interference beyond some region of the lattice and localizes the electron to a finite number of sites.
This phenomenon is known as Aharonov-Bohm ``caging’’ \cite{vidal_aharonov-bohm_1998}.
In the rhombus lattice, Aharonov-Bohm caging occurs at a magnetic field of half a flux quantum through each plaquette, corresponding to a geometric phase of $\pi$.
All three bands become flat and a complete basis of only compact localized states becomes possible (Fig.\,\ref{fig:setup}b).
Accordingly, any initially localized particle will be in a superposition of a finite number of compact localized states and therefore will remain bounded in time. 

Crucially, because of strict boundedness, the key physics of Aharonov-Bohm caging can be accessed through a single plaquette of the rhombus lattice with $\pi$ flux.
In this four-site system, the eigenstates (Fig.\,\ref{fig:setup}d) exhibit similarities to the compact localized states of its lattice counterpart, despite having energies that differ from the continuum limit.
In particular, none of these states extend across an entire plaquette, purely due to the same destructive wavefunction interference effects that give rise to Aharonov-Bohm caging in a lattice.

In this work, we leverage the flexible connectivity of superconducting circuits in a novel manner to realize one of these plaquettes, which we refer to as an Aharonov-Bohm cage.
To allow geometric phase accumulation for microwave photons, which do not inherently respond to magnetic fields due to their neutral charge, we realize a $\pi$-flux synthetic gauge field via a negative tunneling between two sites in the plaquette.
We show how this negative tunneling allows us to turn on caging, by comparison to an analogous device without synthetic field.
Next, we use on-site interactions between particles to demonstrate a bound particle pair escaping the cage.
Finally, we find that unbound particle pairs remain caged in a multi-particle variation of Aharonov-Bohm caging.

\section*{Circuit device}

The sites of our plaquettes are flux-tunable transmon qubits which can be approximated as anharmonic oscillators for microwave photons.
To implement the zero-flux rhombus plaquette, all four qubits are placed in the same orientation as shown in the main image of Fig.\,\ref{fig:setup}e.
Each qubit has a large capacitor comprised of an upper and a lower electrode (shaded green), and tunnel coupling between neighboring qubits is given by capacitance between two of their electrodes.
With upper electrodes capacitively coupled only to lower electrodes of neighboring qubits, the accumulated phase of a single photon traversing this loop is zero.
To generate the $\pi$-flux synthetic field, we rotate one transmon (the rightmost) clockwise by 90 degrees relative to the others.
As shown in Fig.\,\ref{fig:setup}f, in doing so, the (formerly) lower electrode of the right qubit now couples to the lower electrode of the top qubit, while all other electrodes remain coupled as before. 
A single photon traversing this loop then picks up an additional minus sign when hopping between the top and right sites \cite{yanay_two-dimensional_2020}, \emph{i.e.} a geometric phase of $\pi$.
We note that the choice of qubit to rotate and its rotation direction amount to a gauge choice, reflected in where the change in sign of tunneling occurs.

The resulting Hamiltonian for the two devices (in units where the reduced Planck constant $\hbar=1$) is given by:
\begin{align}
\hat{H} = &- \sum_{\langle i,j\rangle} t_{ij} (b^\dagger_i b_j + b_i b^\dagger_j)\nonumber\\
&+ \sum_i \left(\omega_i b^\dagger_i b_i + \frac{U_i}{2} b^\dagger_i b_i (b^\dagger_i b_i - 1)\right) \label{eq:H}
\end{align}
where the summations over sites $i$ and neighboring sites $\langle i, j\rangle$ are taken over the left (L), top (T), bottom (B), and right (R) qubits.
Here, $b_i$ is the bosonic annihilation operator and $\omega_i$ is the frequency of qubit $i$.
For the gauge choice in this work, the tunneling $t_{ij}$ is positive between all qubits in the zero-flux rhombus.
Rotation of the rightmost qubit for the $\pi$-flux rhombus corresponds to changing the sign of tunneling between the right and top qubits ($t_\mathrm{RT} < 0$) while preserving its magnitude.
On-site attractive interactions $U_i<0$ are mediated by the negative qubit anharmonicity of transmon qubits.
Both devices operate in the strongly interacting regime where the ratio of interaction strength to tunneling is approximately 13.5.
Next-nearest neighbor coupling was designed to be negligible at less than $t_{ij} / 70$ and is not included in the model.

We probe each of the two devices via two-tone spectroscopy to confirm that their eigenenergies match the expected energies obtained by exact diagonalization of the corresponding four-site Hamiltonian.
With all four qubits tuned into resonance with each other, we measure transmission through the feedline at the top-qubit resonator frequency while varying the frequency of the pump tone.
Transmission peaks indicate that the resonator frequency has dispersively shifted, which occurs when the pump tone is resonant with a plaquette eigenenergy and excites the top qubit.
The results are shown in Figs.\,\ref{fig:setup}c and d for the zero- and $\pi$-flux rhombi, respectively, and are consistent with calculation.
The measured spectral dependence on one of the qubit frequencies is included as Extended Data Fig.\,\ref{fig:spec}. 

The experiments that follow probe non-equilibrium dynamics and consist of state initialization, time evolution, and readout (Fig.\,\ref{fig:setup}g).
For Hamiltonian time evolution, we set the frequencies of all of the qubits to $\omega_i/2\pi = 4.45$\,GHz. 
At this target frequency, the qubits have depolarization times exceeding 43\,$\mu s$ and nearest-neighbor tunneling amplitudes of approximately $2\pi\!\times\!11.7$\,MHz. 
State initialization and readout are performed via a single feedline (shaded purple in Fig.\,\ref{fig:setup}e), with four half-wave resonators (shaded yellow) dispersively coupled to each qubit. 
In these steps, the frequency of each qubit can be dynamically set between $4.4$ and $5.7$ GHz by varying the current supplied to its individual flux bias line (shaded blue, Fig.\,\ref{fig:setup}e).
A full table of device parameters can be found in Supplementary Information Tabs.\,\ref{tab:params} and \ref{tab:t}.

\section*{Caging of a single particle}

Aharonov-Bohm caging can be made apparent through a single-particle quantum walk \cite{abilio_magnetic_1999, naud_aharonovbohm_2002, mukherjee_experimental_2018, hung_quantum_2021, li_aharonov-bohm_2022}.
Much like the walk in a one-dimensional chain \cite{karski_quantum_2009, preiss_strongly_2015, yan_strongly_2019, gong_quantum_2021, karamlou_quantum_2022}, the ensuing dynamics result from wavefunction interference.
Here, destructive interference bounds the quantum walk in an Aharonov-Bohm cage, while constructive interference allows a fully delocalized walk over all four sites of the plaquette with zero flux.

\begin{figure}[t!]
    \centering
     \includegraphics[width=\columnwidth]{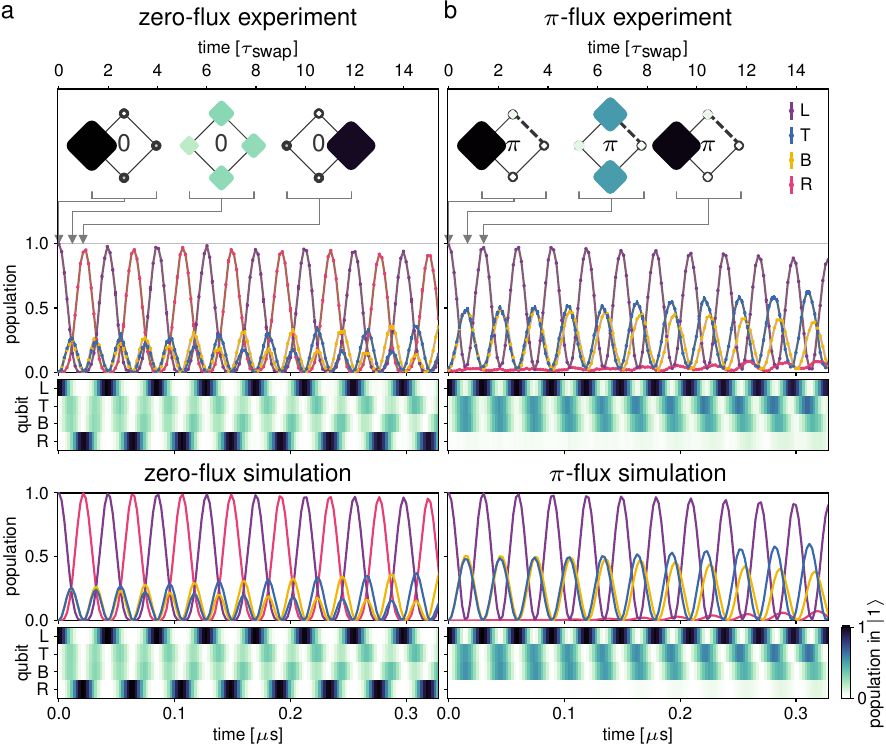}
    \caption{Aharonov-Bohm caging of a single particle. \textbf{(a)} Average qubit populations in the first excited state $|1\rangle$ as a function of time after quench into resonance (upper), for the zero-flux plaquette. Appreciable populations are observed on all four sites, even at short times, indicating a fully delocalized walk. Insets show average qubit populations in the initial state and after approximately $1/2$ and $1$ swap times.
    \textbf{(b)} In the $\pi$-flux plaquette, the single particle remains bounded to three of the four sites as a result of Aharonov-Bohm caging. Here insets correspond to approximately $0$, $1/\sqrt{2}$, and $\sqrt{2}$ swap times.
    In both (a) and (b), simulated time evolution with zero on-site disorder is shown in the lower half. 
    Population time series are also represented as color plots for visibility and use the colorbar in the lower right of the figure. 
    Duplicate time axes are included to indicate time in units of a photon swap time, $\tau_{\mathrm{swap}}$ (see main text for details).
    For data in this and subsequent figures, error bars indicate 95\% Clopper-Pearson confidence intervals for 3000 measurements per data point after correcting for readout errors and may be smaller than the markers. Adjacent markers are connected with a darker-shade solid line to guide the eye.}
    \label{fig:1particle}
\end{figure}

The experiment begins with initialization of the leftmost qubit, detuned from the target qubit frequency, in the first excited state (see Fig.\,\ref{fig:setup}d).
This is equivalent to placing a single particle (microwave photon) on that site.
We then quench the left qubit into resonance with the other three qubits at the target qubit frequency.
After evolution under the Hamiltonian (Eq.\,\ref{eq:H}) for a variable time, we diabatically detune and measure the state of one qubit.
Repeated measurements across all qubits then provide access to the average site-resolved occupation of all four sites.
The results for zero and $\pi$ flux, corrected for readout errors, are shown in Fig.\,\ref{fig:1particle}a and b (upper), respectively.
Duplicate time axes are included in units of $\tau_{\mathrm{swap}} = 2\pi / 4\bar{t}$: the time it takes for a photon to swap between two qubits with tunnelling rate $\bar{t}$, where $\bar{t} = 2\pi\!\times\!11.75 \,(2\pi\!\times\!11.66)$\,MHz is the average rate for the zero-flux ($\pi$-flux) device.
Simulation results with zero on-site disorder and independently measured tunneling and decoherence are plotted below.

We find a bounded quantum walk in the $\pi$-flux rhombus, indicating successful Aharonov-Bohm caging.
Within the first swap time after quench, the particle begins to delocalize over the top and bottom sites in the plaquette regardless of flux.
Shortly thereafter, however, the walks deviate.
In the zero-flux rhombus, the top and bottom paths constructively interfere and result in a periodic walk across all four qubits, as demonstrated by the insets in Fig.\,\ref{fig:1particle}a, upper.
This does not occur for the $\pi$-flux rhombus: interference is destructive on the rightmost site, suppressing particle population below 5\% for over 9 swap times (Fig.\,\ref{fig:1particle}b).
Consequently, the particle remains bounded to the left, top, and bottom sites (insets in Fig.\,\ref{fig:1particle}b, upper) as expected for caging.

The agreement between experimental measurement and simulation indicates that our control over qubit frequencies is consistent with achieving negligible on-site disorder and sufficient to realize caging.
The slight deviation from ideal caging is instead limited predominantly due to disorder in the coupling strengths between qubits, which amounts to a spread of approximately 2.5\% for the zero-flux device and 5.6\% for the $\pi$-flux device.
This coupling-strength disorder results in coherent beating which is highlighted at long timescales (Extended Data Fig.\,\ref{fig:longtimes}).

\section*{Doublons escaping the cage}

We next consider the walk of two particles initialized on the same site, referred to as a doublon.
These doublon states span a subspace of the entire two-particle Hilbert space; the complement is spanned by two-particle states where the particles sit on different sites, which we will refer to as ``particle-particle'' Fock states.
In the presence of strong interactions, the doublon states are energetically well-separated from the particle-particle states. 
In this subspace, doublons hop as a bound pair and experience a gauge field that is twice that of a single particle. 
The Aharonov-Bohm cage for a single particle, then, has no effect on a strongly interacting doublon.

We access the strongly interacting regime with our devices by designing interaction energies in excess of tunneling amplitudes by an order of magnitude.
The doublon walk begins with two photons on the left site and continues with time evolution under the Hamiltonian as before, where on-site energies are now tuned to ensure that the second excited states of all qubits are on resonance.
After a variable time, we measure the population of the second excited state and extract readout-error-corrected average doublon occupations of all sites.

\begin{figure}[t!]
    \centering
     \includegraphics[width=\columnwidth]{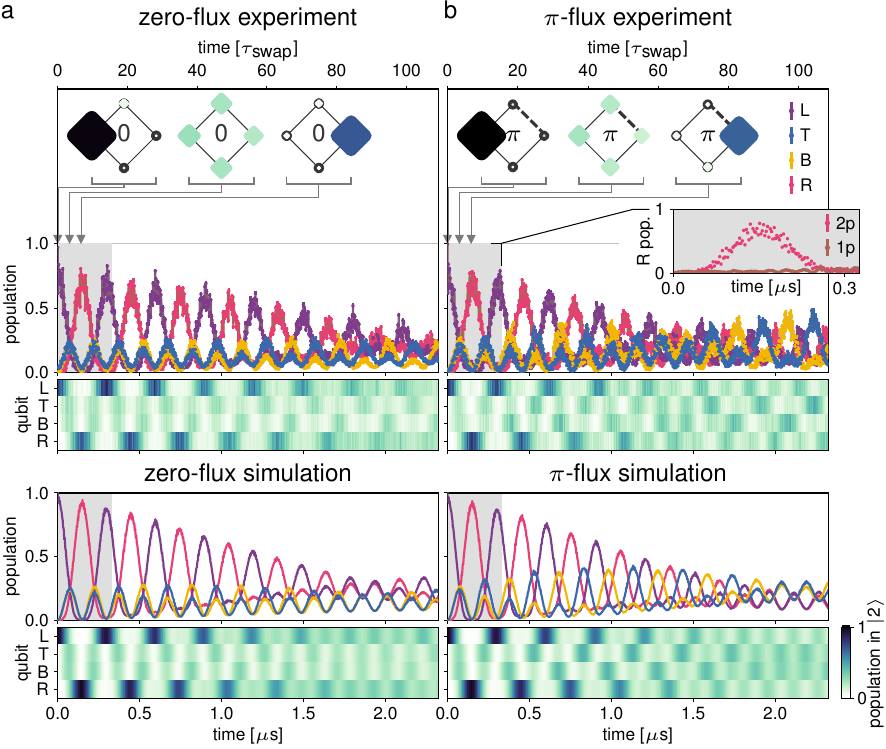}
    \caption{Escape of an interacting photon pair. \textbf{(a)} Average qubit populations in the second excited state $|2\rangle$ as a function of time after quench into resonance (upper), for the zero-flux plaquette. The walk remains fully delocalized as with a single particle, but at reduced timescales.
    \textbf{(b)} Much like in the zero-flux plaquette, the doublon walk extends across the full $\pi$-flux plaquette. (Inset plot) Direct comparison of right-qubit populations (R pop.) for doublon (2p, $|2\rangle$ population) and single-particle (1p, $|1\rangle$ population) walks in the $\pi$-flux plaquette, highlighting the doublon escape in contrast to single-particle caging.
    In both (a) and (b), inset plaquettes show average qubit populations in the initial state and after approximately $1/2$ and $1$ doublon-renormalized swap times. Simulated time evolution with zero on-site disorder, but including measured tunneling disorder and qubit decoherence, is shown in the lower half.
    The shaded gray regions indicate the total evolution time shown in Fig.\,\ref{fig:1particle}.
    Population time series are also shown as color plots for visibility and use the colorbar in the lower right of the figure.}
    \label{fig:2particle}
\end{figure}

Doublons in the zero-flux rhombus delocalize in a manner qualitatively similar to single photons (Fig.\,\ref{fig:2particle}a).
Doublon dynamics in the $\pi$-flux rhombus, however, markedly differ from single-photon caging in that the doublon does not remain bounded away from the rightmost site (Fig.\,\ref{fig:2particle}b inset plot).
Rather, the doublon fully explores the plaquette almost as if it were in the zero-flux plaquette, \emph{i.e.} it escapes this Aharonov-Bohm cage. 

We note that in both plaquettes, the doublon tunneling amplitude is reduced by almost an order of magnitude compared to the single-particle tunneling.
These timescales result from doublons hopping via a second-order process, with rate given by $2t^2/U$ for bare tunneling $t$ and negative anharmonicity $U$.
This effective tunneling provides another way of understanding the doublon escape: all four tunneling rates $2t_{ij}^2/U$ have the same sign regardless of the sign of $t_{ij}$, yielding zero flux.
Additionally, the doublon-population oscillation amplitudes decrease over time, largely due to the weak hybridization between the doublon and particle-particle states, leading to coherent beating of the measured doublon population (see Supplementary Information for additional discussion).

\section*{Particle-Particle Fock-State Caging}

Finally, we investigate the time dynamics of particle-particle Fock states and find a variation of Aharonov-Bohm caging.
These states, where two particles sit on different sites and are not bound by interactions, comprise the second and remaining sector of two-particle states.
In fact, strong interactions energetically separate these states from the doublons, making the individual particles behave approximately as hard-core bosons.

\begin{figure}[t!]
    \centering
     \includegraphics[width=\columnwidth]{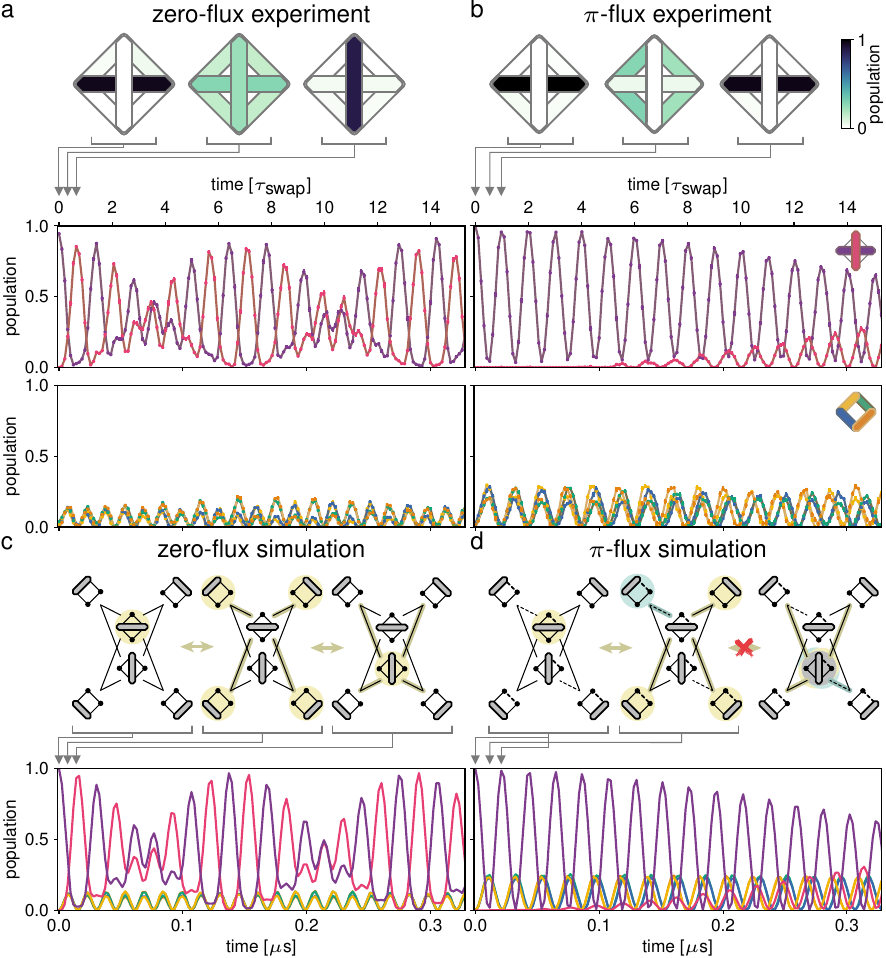}
    \caption{Particle-particle Fock-state caging \textbf{(a)} Average populations of each of the six particle-particle Fock states for variable time after quench into resonance. The initial state (left inset) consists of one photon on the left site and one photon on the right. Other insets show Fock-state populations at approximately $1/(2\sqrt{2})$ and $1/\sqrt{2}$ swap times. 
    In insets, particle-particle Fock-state populations are shown via the corresponding segment that extends across two sites of the plaquette, \emph{e.g.} the horizontal segment corresponds to $|\mathrm{LR}\rangle$.  
    The top panel plots the population of particle-particle Fock states of next-nearest-neighbor sites, while the middle panel shows nearest-neighbor particle-particle Fock state population. \textbf{(b)} Same, but for the $\pi$-flux plaquette. Insets show Fock state populations of the initial state and at approximately $1/2$ and $1$ swap times. \textbf{(c)} Simulated time evolution with zero flux and with zero on-site disorder for all six particle-particle Fock state combinations, incorporating measured tunneling disorder and qubit decoherence. Hamiltonian adjacency graphs depict the particle-particle dynamics. \textbf{(d)} Same as c, but for the $\pi$-flux plauqette. Here, the Hamiltonian adjacency graph describes a caging-like effect where tunneling of opposite sign and states with relative phase $\pi$ (teal shading) cause destructive interference and $|\mathrm{TB}\rangle$ remains unpopulated.}
\label{fig:interactingwalk}
\end{figure}

For these experiments, we initialize one photon on the left site and one photon on the right (the $|\mathrm{LR}\rangle$ state).
After time evolution, we perform simultaneous readout on all qubit pairs to extract average particle-particle Fock-state populations at various points in time and plot the results in Fig.\,\ref{fig:interactingwalk}a and b. 
The particle-particle dynamics in the zero-flux plaquette are dominated by full-contrast swaps between the $|\mathrm{LR}\rangle$ and $|\mathrm{TB}\rangle$ states.
In contrast, we find a bounded walk in the $\pi$-flux plaquette not in real space, but instead characterized by suppressed population transfer to the $|\mathrm{TB}\rangle$ Fock state. 
We refer to this variation of caging as ``Fock-state caging".

This bounded phenomenon is a direct consequence of Aharonov-Bohm caging extended to multi-particle sectors, seen through the corresponding Hamiltonian adjacency graphs in Figs.\,\ref{fig:interactingwalk}c and d.
In these graphs, each node represents a particle-particle Fock state; doublon states are omitted in this hard-core limit.
Two nodes are connected by edges if the Fock states are tunnel-coupled, and edges are weighted according to the coupling sign and magnitude.
In this configuration space, time evolution of the $|\mathrm{LR}\rangle$ state corresponds to a single-particle walk on the adjacency graph beginning on the $|\mathrm{LR}\rangle$ node.
Without a synthetic field, delocalization proceeds across the entire particle-particle subspace.
Whereas, with $\pi$ flux, destructive interference much like in Aharonov-Bohm caging prohibits evolution to the $|\mathrm{TB}\rangle$ state.
In other words, Fock-state caging results from the interplay of geometric phase accumulation and Fock-state configurational-space topology, rather than real-space lattice topology as in Aharonov-Bohm caging.
In multi-plaquette systems and the infinite lattice itself, this Fock-state caging may give rise to an interacting two-particle real-space caging effect \cite{santos_methods_2020}; see Extended Data Fig.\,\ref{fig:tripleplaquette} for supporting evidence from simulations on a chain of three plaquettes.

The oscillating populations between the $|\mathrm{LR}\rangle$ and $|\mathrm{TB}\rangle$ states in the zero-flux rhombus display a beating effect. 
This coherent beating can be attributed to weak hybridization between the doublon and particle-particle states as in the doublon walk, reflecting a slight deviation of our device parameters from the hard-core regime (see Supplementary Information for details). 
Interestingly, the $\pi$-flux rhombus does not host beating of these populations beyond what can be attributed to tunneling-rate disorder. 
The absence of beating suggests that the Fock-state caging effects are independent of interaction strength. 
In the regime of finite interactions, doublon states must be included in the adjacency graph; even so, we find that the resulting configuration-space topology maintains the complete destructive interference necessary for Fock-state caging.

\section*{Outlook}
Our work experimentally demonstrates the Aharonov-Bohm caging of a single particle and the subsequent escape of deeply bound doublons from the cage.
We realize this cage in a rhombus plaquette by engineering a negative tunnel coupling between exactly two of the transmon qubits to thread a $\pi$-flux synthetic magnetic field through the plaquette.
Additionally, we consider the correlated dynamics of two photons initialized on opposite sides of the plaquette and find a type of caging not in real space, but in the configurational space of particle-particle Fock states.
These results establish a critical building block for studies of all-bands-flat lattices with strong interactions \cite{vidal_interaction_2000, kolovsky_conductance_2023} and inform future studies in other flat-band lattices with superconducting circuits, both in one- \cite{ deng_superconducting_2016, gneiting_lifetime_2018} and two-dimensions.
For example, larger system sizes and increased coherence may enable the characterization of disorder and thermalization in a flat band \cite{chalker_anderson_2010, longhi_inverse_2021}.
Combined with dissipative stabilizers \cite{ma_dissipatively_2019}, equilibrium states may also be within reach \cite{biondi_incompressible_2015, katsura_mott_2021}.
Finally, other platforms such as ultracold atoms can be used \cite{flannigan_enhanced_2020, junemann_exploring_2017} to study fermions \cite{kobayashi_superconductivity_2016} and nonitinerant spin models \cite{mcclarty_disorder-free_2020}, implement longer-range interactions in these flat-band lattices \cite{roy_interplay_2020, salerno_interaction-induced_2020, tilleke_nearest_2020, danieli_many-body_2020, khare_localized_2021}, or measure transport between two reservoirs \cite{pyykkonen_flat_2021}.

\vspace{10pt}

\begin{acknowledgements}
    $^*$These authors contributed equally to this work.
    
    \textbf{Acknowledgements}
    We thank Sara Sussman and Sho Uemura for assistance on QICK software development and Lawrence Cheuk, Anjali Premkumar, and Rhine Samajdar for discussions.
    We acknowledge support from the NSF Quantum Leap Challenge Institute for Robust Quantum Simulation 2120757, the ARO MURI W911NF-15-1-0397, and the Princeton Center for Complex Materials NSF DMR-1420541.
    J.G.C.M. acknowledges additional support from NSF GRFP DGE-2039656.
    This research made use of the Micro and Nano Fabrication Center at
    Princeton University.
	
    \textbf{Contributions}
    C.S.C. conceived and designed the experiment with J.G.C.M.
    J.G.C.M. designed the devices with assistance from B.M.S., and fabricated the devices.
    J.G.C.M. performed the experiments with C.S.C. and together analyzed the data.
    All authors contributed extensively to the interpretation of data, writing of the manuscript, and discussions.    
	
    \textbf{Corresponding author} Correspondence and requests for materials should be addressed to A.A.H. 
    
\noindent{(aahouck@princeton.edu).}
	
    \textbf{Competing interests} The authors declare no competing interests. 
\end{acknowledgements}

\section*{Methods}

\subsection{Device fabrication}
Both 10\,mm $\times$ 10\,mm devices were patterned on a 200\,nm-thick layer of tantalum deposited on a 530$\,\mu $m thick sapphire wafer. 
Direct-write optical lithography was used to define the base layer pattern and the features were etched with a Chlorine-based dry etch.
Josephson junctions were patterned using electron-beam lithography and the Al-AlO$_x$-Al layers were deposited using double-angle evaporation in a Plassys MEB 550S. 
The first arm was 20\,nm thick, the AlO$_x$ was grown using a 85/15 Ar/O$_x$ mix at 200\,mBar for 40 minutes, and the second aluminum layer was 70\,nm thick. 
Both devices were packaged in QDevil QCage aluminum sample holders. 
Wirebonds were used to short the ground planes on opposite sides of coplanar waveguides, ground the metal in the middle of the four qubits, and bond signal traces to bondpads.   

\subsection{Wiring \& setup}
Devices were measured sequentially in a Bluefors LD250 at a base temperature of 11 mK. 
All time-domain control and measurements were performed using the QICK-controlled Xilinx RFSoC ZCU216 \cite{stefanazzi_qick_2022}, including resonator and qubit drives, qubit fast-flux pulses, and single- and simultaneous multiple-cavity readout. 
DC-flux offset for each qubit was controlled using Yokogawa low-noise voltage sources, combined with a fast-flux line in the mixing chamber, and sent to the qubit via on-chip flux lines.

The qubits were dispersively read out and driven through individual half-wave resonators capacitively coupled to a single transmission line. 
Cavity readout tones and qubit drive pulses were combined in a splitter at room temperature. 
The output signal was amplified using a high-electron-mobility transistor (HEMT) amplifier at 4K and two amplifiers at room temperature.
The full wiring setup can be found in Supplementary Information Fig.\,\ref{fig:fridge}.

\subsection{Experimental sequence}
For all experiments, DC-flux voltages were set such that in the absence of fast-flux pulses, all four qubits were approximately on resonance with one another. 
State initialization, time evolution, and readout were then performed using fast-flux pulses that control the qubit frequencies within approximately one nanosecond. 
State initialization was done by detuning only the qubits with initial-state occupation by approximately $20t$ and applying individual $\pi$ pulses. 
Time evolution proceeded by rapidly tuning the qubits onto resonance and waiting for a variable amount of time, after which qubits were detuned by $20t$ for single-shot readout.
For the one-photon and doublon walks, we detuned and measured a single qubit to reduce effects of frequency crowding and unwanted population transfer. 
For the particle-particle walk, two qubits were simultaneously detuned and read out to measure correlations. 
To capture the populations of all qubit and qubit pairs, we repeat each experimental run and vary which qubit(s) are measured. 

All readout consists of single-shot measurements using 3000 shots. 
Readout parameters and thresholds were optimized to discriminate between the ground and first excited states for measurements of the first-excited-state population, and between the first and second excited states for measurements of the second-excited-state population.
Averaging time for readout was 2.5\,$\mu$s.

\subsection{DC-flux crosstalk calibration}
We calibrated and compensated for DC-flux crosstalk as in  \cite{braumuller_probing_2022}.
We find a non-negligible crosstalk of around 2\%.
The full crosstalk matrix for the $\pi$-flux device is shown in Supplementary Information Fig.\,\ref{fig:xc}; the crosstalk matrix for the zero-flux device is similar.

\subsection{Fast-flux calibration}
The fast-flux pulses used to tune qubit frequencies during the experimental sequence ideally have step-function responses.
We calibrated for long-time deviations from the ideal step pulse as in  \cite{ma_dissipatively_2019} and short-time deviations as in  \cite{braumuller_probing_2022}.
One and two-photon oscillations between two qubits before and after these compensations are plotted in Supplementary Information Fig.\,\ref{fig:FF}.

\subsection{Qubit tuning into resonance}
The four qubits were first biased approximately into resonance using DC-flux currents. 
To compensate for frequency disorder, we fine-tuned the qubit frequencies using fast-flux pulses through a heuristic algoritm.
We first set the bottom qubit to a frequency near its sweet spot and varied the step amplitude of a fast-flux pulse to bring the left qubit into resonance, probed via a single-particle two-qubit walk.
Next we brought the top qubit into resonance, probed via the dynamics of a single particle initialized on the left qubit and optimized for balanced measurement probabilities on the top and bottom qubits.
Finally, we brought the right qubit into resonance, verified through comparison with simulated plaquette dynamics.

\subsection{Hamiltonian estimation \& simulation}
All of the parameters used in the Hamiltonian time-domain simulations were independently measured.
To determine tunneling amplitudes between neighboring qubit pairs, we performed a single-particle two-qubit walk with the nonparticipating qubits detuned by at least $250$ MHz.
This walk was also used to determine approximate $T_2$ dephasing times for the coupled plaquette systems, which have a reduced flux dispersion compared to that of an individual qubit \cite{guo_observation_2021}.
The $T_1$ coherence time was individually measured at the target frequency.
Simulations use the QuTiP package \cite{johansson_qutip_2013}.

\setcounter{section}{0}
\setcounter{subsection}{0}
\setcounter{figure}{0}
\setcounter{equation}{0}

\makeatletter
     \@addtoreset{figure}{section}
\makeatother

\renewcommand{\thesection}{\arabic{section}}
\renewcommand{\thesubsection}{\thesection.\arabic{subsection}}
\renewcommand{\thesubsubsection}{\thesubsection.\arabic{subsubsection}}

\makeatletter
\renewcommand{\p@subsection}{}
\renewcommand{\p@subsubsection}{}
\makeatother
\renewcommand{\numberline}[1]{#1~}

\makeatletter
\renewcommand{\thefigure}{S\@arabic\c@figure}
\renewcommand{\thetable}{S\@arabic\c@table}
\makeatother
\renewcommand{\theequation}{S\arabic{equation}}

\newpage
\onecolumngrid
\section*{Extended Data}

\begin{figure}[H]
    \centering
     \includegraphics{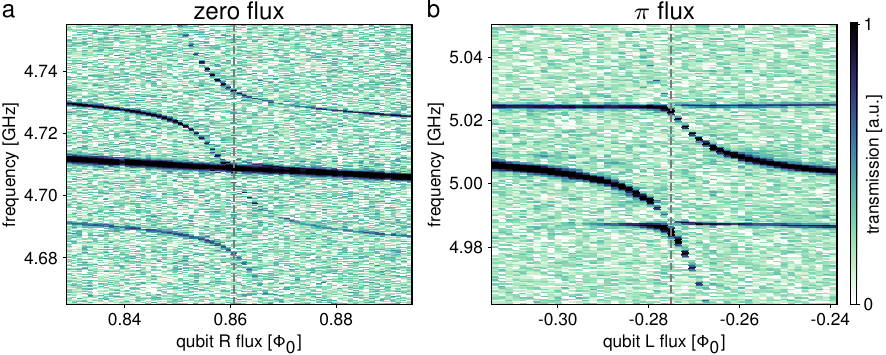}
    \caption{Two-tone spectroscopy data for the \textbf{(a)} zero- and \textbf{(b)} $\pi$-flux rhombus devices. In both, the DC voltage on one of the qubit flux lines is varied, keeping the other three qubit frequencies in resonance with one another. Then, the pump tone frequency is varied while the transmission at the top qubit readout resonator frequency $\omega_{r,0}$ is monitored. Increased transmission corresponds to the pump tone being on resonance with one of the four-qubit-system eigenfrequencies because it shifts the qubit readout resonator frequency away from the probe tone frequency. The gray dashed lines indicate the flux values taken for the spectroscopy data shown in Fig.\,\ref{fig:setup}c, d of the main text.}
    \label{fig:spec}
\end{figure}

\begin{figure}[H]
    \centering
    \includegraphics[width=\textwidth]{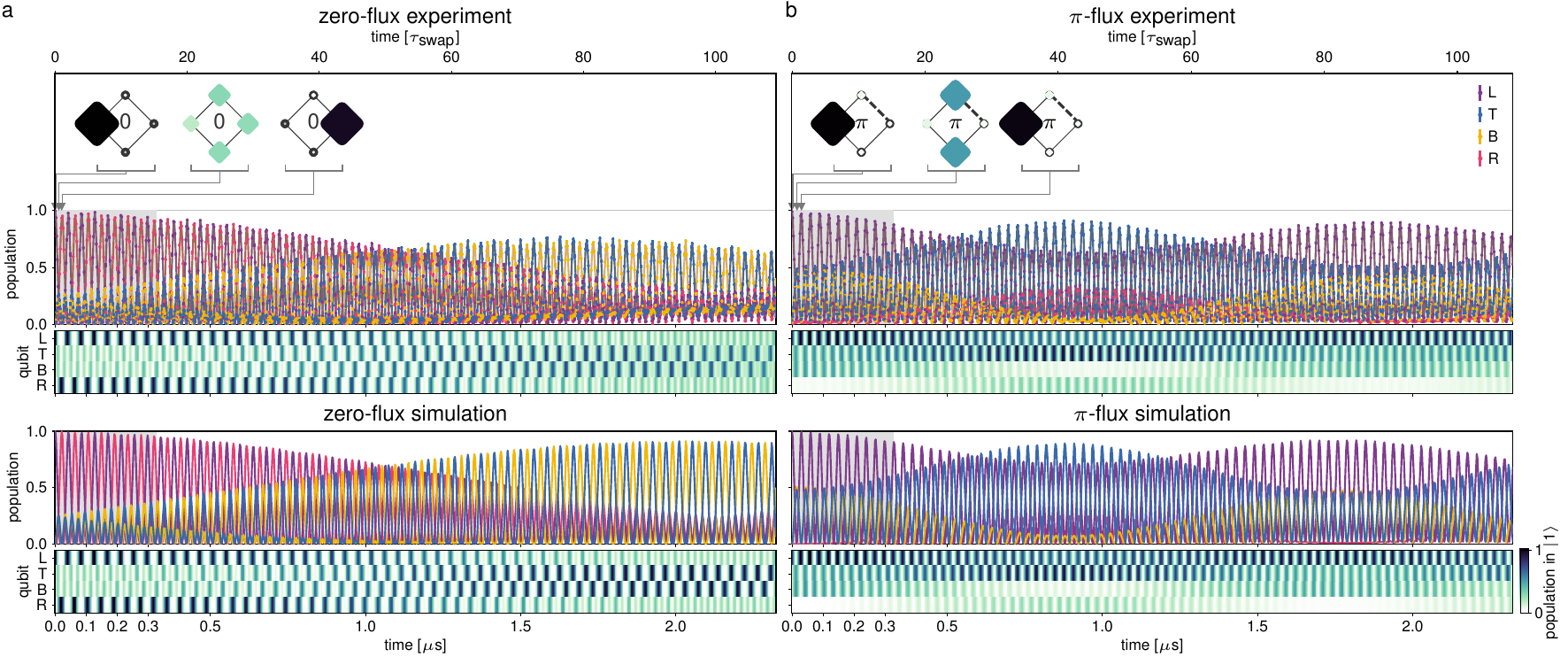}
    \caption{One-photon walk out to longer times for the \textbf{(a)} zero- and \textbf{(b)} $\pi$-flux rhombi. The datasets shown in main text Fig.\,\ref{fig:1particle} are displayed here out to longer times, with the gray rectangle indicating the time range of main text Fig.\,\ref{fig:1particle}. Population time series are also represented as color plots for visibility and use the colorbar in the lower right of the figure. 
    The long-timescale dynamics seen here, namely photon population on the right qubit, results from inhomogenous nearest-neighbor tunneling amplitudes as can be seen in the simulation (lower half of plot).
    Error bars indicate 95\% Clopper-Pearson confidence intervals for 3000 measurements per data point and may be smaller than the markers. Adjacent markers are connected with a darker-shade solid line for visibility. Adjacent markers are connected with a darker-shade solid line to guide the eye. Insets come from Fig.\,\ref{fig:1particle} and are included for reference.}
    \label{fig:longtimes}
\end{figure}

\begin{figure}[H]
    \centering
    \begin{minipage}{0.50\textwidth}
     \includegraphics[width=\textwidth]{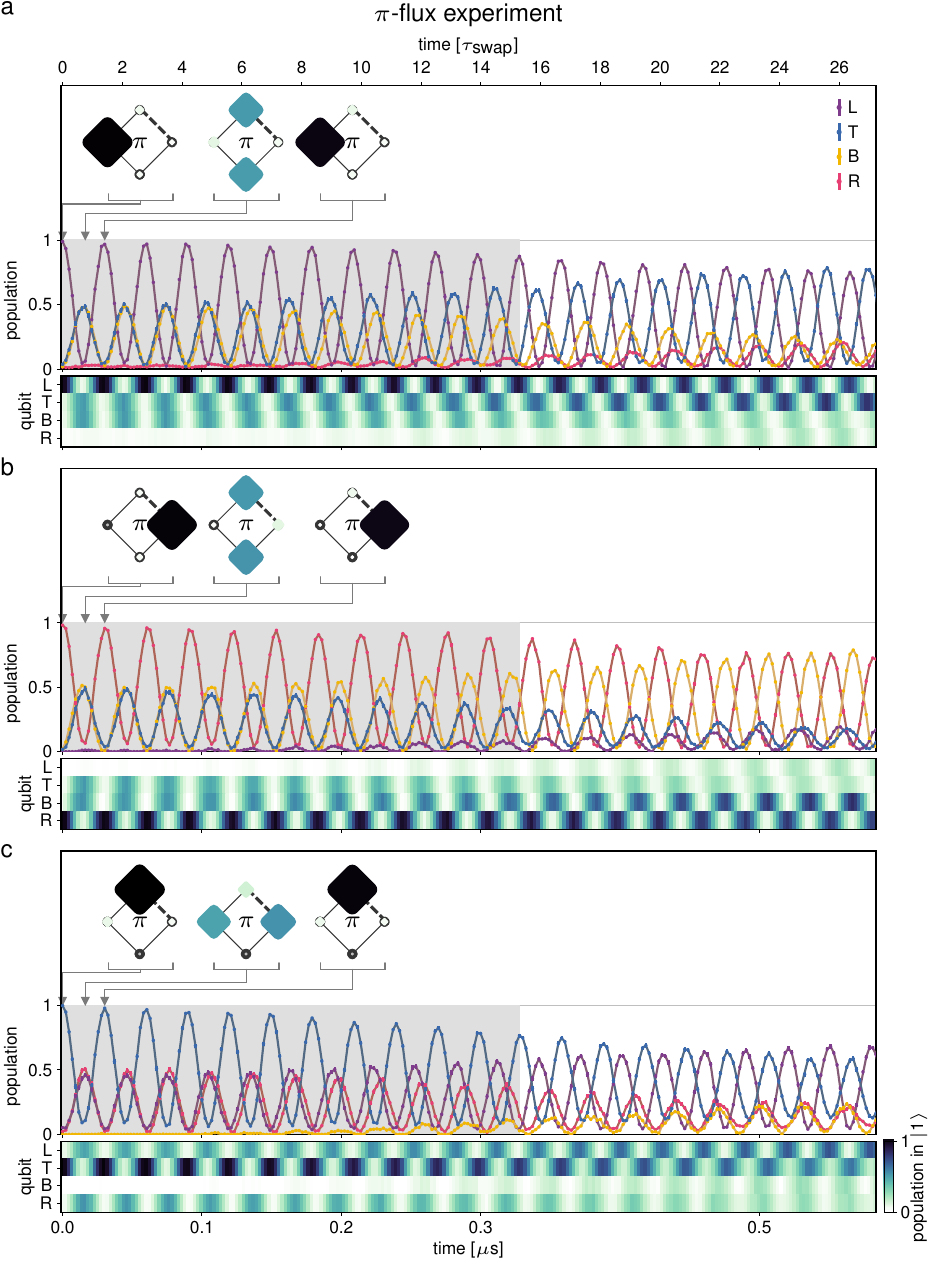}
     \end{minipage}
     \begin{minipage}{0.49\textwidth}
    \caption{One-photon walk with varying initial state for $\pi$-flux plaquette. \textbf{(a)} Initializing photon on left qubit, as in main text Fig.\,\ref{fig:1particle} and Extended Data Fig.\,\ref{fig:longtimes}. \textbf{(b)} Initializing photon on right qubit. The dynamics are qualitatively similar to (a), as expected, with left-qubit population suppression below 5\% out to approximately 11.7 swap times. \textbf{(c)} Initializing photon on upper qubit. Again, dynamics are qualitatively similar to (a), with bottom-qubit population suppression below 5\% out to approximately 11.8 swap times.
    In (a-c), population time series are also represented as color plots for visibility and use the colorbar in the lower right of the figure. 
    Error bars indicate 95\% Clopper-Pearson confidence intervals for 3000 measurements per data point and may be smaller than the markers. Adjacent markers are connected with a darker-shade solid line for visibility. Adjacent markers are connected with a darker-shade solid line to guide the eye. Inset plaquettes show average qubit populations at approximately $0$, $1/\sqrt{2}$, and $\sqrt{2}$ swap times. }
    \label{fig:walkdir}
    \end{minipage}
\end{figure}

\vspace{30pt}
\begin{figure}[H]
    \centering
     \includegraphics{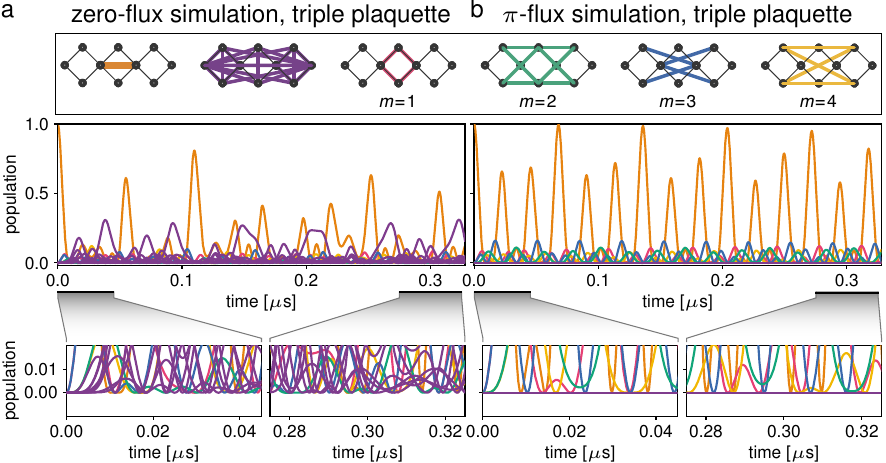}
    \caption{Particle-particle walk on the triple plaquette with $|\mathrm{LR}\rangle$-like initial state. Legend applies to all plots and groups multiple Fock-state populations under the same color for visibility, however each population is plotted as separate trace. The legend indicates the initial state population, all particle-particle Fock-state populations which exhibit zero population at all times in the $\pi$-flux lattice, and the remaining Fock-state populations grouped by particles on sites which are distance $m$ sites from each other.  \textbf{(a)} Zero synthetic magnetic field. The interacting walk covers all particle-particle Fock states over time.
    \textbf{(b)} With the $\pi$-flux field, the interacting walk never evolves to exhibit population on the analogous $|\mathrm{TB}\rangle$-like state (purple). Furthermore, this photon-photon state appears to remain bounded; the leftmost and rightmost sites of the system remain unpopulated at all points in time (purple). Simulations use the average tunnel couplings and interactions energies for each device, respectively, and omit disorder and dephasing. Insets zoom in around the zero population point at short and long times.}
    \label{fig:tripleplaquette}
\end{figure}

\newpage
\onecolumngrid
\begin{center} \begin{Large}
    Supplementary Materials for\\\vspace{5pt} \textbf{Flat-band localization and interaction-induced delocalization of photons}
\end{Large} \end{center}

\section{Device parameters}

\begin{table}[H]
\centering
    \setstretch{1.5}
    \scalebox{1}{
    \begin{tabular}{r|r r r r|r r r r}
	& \multicolumn{4}{c}{zero flux} & \multicolumn{4}{c}{$\pi$ flux} \\
        qubit & left & top & bottom & right & left & top & bottom & right\\
    \hline
        $E_{J,\mathrm{max}} /2\pi$ (GHz) & 26.80(1) & 25.85(1) & 27.69(1) & 26.32(1) & 27.58(1) & 27.01(1) & 27.72(1) & 27.04(1)\\
        $d$ & 0.5954(3) & 0.5962(4) & 0.5922(4) & 0.5897(2) & 0.5897(2) & 0.5941(3) & 0.5770(3) & 0.6029(2)\\
        $E_C /2\pi$ (MHz) & 158.10(6) & 157.90(5) & 157.80(5) & 157.26(4) & 157.30(2) & 157.56(3) & 158.91(5) & 158.40(7)\\
        $\omega_q /2\pi$ (MHz) & \multicolumn{4}{c|}{4450} & \multicolumn{4}{c}{4450}\\
        $\omega_{r,0} /2\pi$ (MHz) & 6962.62(1) & 7033.80(1) & 7104.82(1) & 7230.25(1) & 6952.62(1) & 7055.99(1) & 7117.79(1) & 7250.22(1)\\
        $\kappa /2\pi$ (MHz) & 0.84(1) & 0.59(1) & 0.81(1) & 0.32(1) & 0.86(1) & 0.54(1) & 0.80(1) & 0.37(1)\\
        $g /2\pi$ (MHz) & 64.73(18) & 64.10(4) & 67.16(9) & 64.63(3) & 60.27(13) & 67.86(10) & 68.40(13) & 67.20(9)\\
        $T_{1, 01}$ (us) &42.4(8)&42.2(1.0)&44.1(1.0)&48.5(1.0)& 50.6(9) & 43.6(7) & 50.4(1.6) & 44.0(1.0)\\
        $T_{2, 01}^*$ (us) &2.95(13)&2.11(9)&3.44(15)&3.08(13)& 2.94(10) & 1.82(7) & 1.96(10) & 3.47(18)\\
        $T_{1, 12}$ (us) &32.6(1.3)&22.9(1.2)&32.8(0.9)&19.2(5)& 27.3(5) & 28.1(6) & 27.9(8) & 27.8(6)\\
        $T_{2, 02}^*$ (us) &1.41(12)&0.89(8)&1.56(10)&1.30(14)& 1.59(14) & 0.90(4) & 1.05(10) & 1.84(10)
    \end{tabular}
    }
    \setstretch{1}
    \caption{Device parameters for the zero- and $\pi$-flux devices presented in this work. $E_{J,\mathrm{max}}$ is the maximum effective Josephson energy, $d$ is the junction asymmetry, and $E_C$ is the capacitor energy which roughly determines the interaction energy $U \approx -E_C$. $\omega_q / 2\pi$ is the target frequency that all four qubits are set to during the evolution portion of the experiment. $\omega_{r,0}/ 2\pi$ is the frequency of the dispersively coupled resonator when the qubit is in the ground state, with its photon loss rate $\kappa$ and coupling to qubit $g$. $T_{1, 01}$ and $T_{2, 01}^*$ are the qubit decay and Ramsey dephasing times between the ground and first excited state at the target qubit frequency 4.45\,GHz, while $T_{1, 12}$ is the qubit decay between the first and second excited states and $T_{2, 02}^*$ is the dephasing time between the second excited state and the ground state. Uncertainties are in parentheses and reflect one standard deviation errors on the fits.}
    \label{tab:params}
\end{table}

Tab.\,\ref{tab:params} summarizes the device parameters for our zero- and $\pi$-flux rhombus circuits; however we note that these exact values are not necessary to observe the results presented in the main text.
Our independently measured tunnel couplings are in Tab.\,\ref{tab:t}.
All qubits are flux-tunable with loop sizes of approximately $45 \mu\mathrm{m} \times 7 \mu\mathrm{m}$, with approximately 2.2 mA corresponding to one flux quantum through the loop.

\begin{table}[H]
\centering
    \setstretch{1.5}
    \footnotesize
    \begin{tabular}{r|c c}
        tunneling & zero flux & $\pi$ flux\\
    \hline
        $t_{\mathrm{LT}} / 2\pi$ (MHz) & 11.879(1) & 11.781(1)\\
        $t_{\mathrm{LB}} / 2\pi$ (MHz) & 11.792(1) & 11.884(1)\\
        $t_{\mathrm{RT}} / 2\pi$ (MHz) & 11.587(1) & 11.736(1)\\
        $t_{\mathrm{RB}} / 2\pi$ (MHz) & 11.734(1) & 11.238(1)\\
    \end{tabular}
    \setstretch{1}
    \caption{Tunneling amplitudes of the devices presented in this work, measured at the target qubit frequency. The labels L, T, B, R correspond to the left, top, bottom, and right qubits, respectively.}
    \label{tab:t}
\end{table}

The qubit target frequency was chosen to reduce frequency sensitivity to flux bias and optimize qubit coherence.
The former consideration merits proximity to a qubit sweet spot, while the second requires avoiding frequency resonances with lossy two-level systems in the device material that are physically proximal to one of the four qubits.
For both devices, we select a qubit target frequency of 4.45\,GHz.

\section{Wiring \& setup}

\begin{figure}
    \centering
    \includegraphics{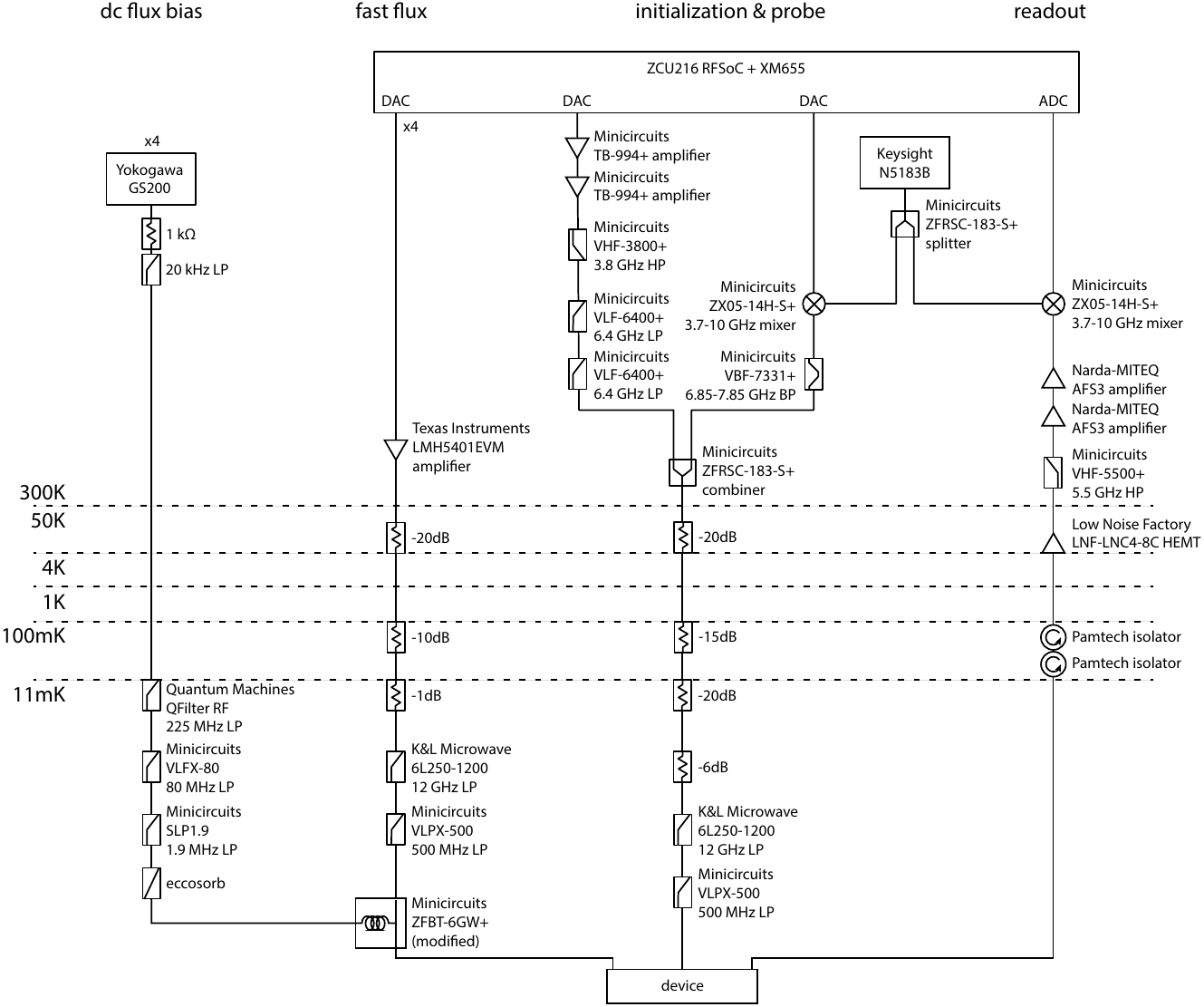}
    \caption{Control and measurement wiring setup.}
    \label{fig:fridge}
\end{figure}

The wiring and measurement setup is in Fig.\,\ref{fig:fridge}, discussed in the Methods, and is generally similar to the setup used in similar works.
We highlight the use of the Xilinx RFSoC ZCU216, with software control through the QICK package \cite{stefanazzi_qick_2022}, for arbitrary waveform generation of the fast-flux pulses, qubit drives, and qubit readout probe.
This board is also used for signal digitization, greatly simplifying the microwave hardware requirements for the setup.
We elaborate also on the modified bias-tee to combine the DC and fast-flux lines in the mixing chamber; here, the capacitor was removed and replaced with a short.

\section{Device flux calibrations}

\begin{figure}
    \centering
    \begin{minipage}{0.35\textwidth}
    \includegraphics{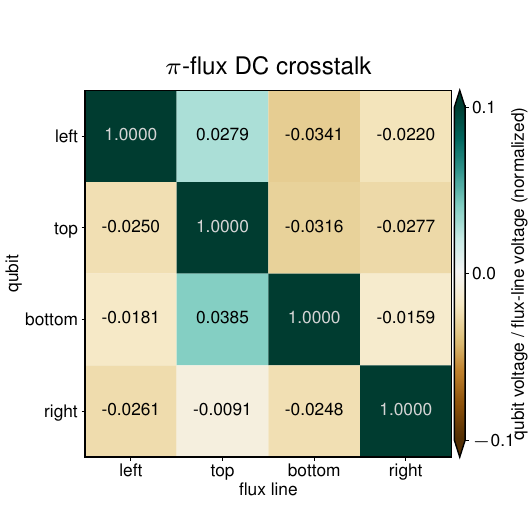}
    \end{minipage}
    \begin{minipage}{0.6\textwidth}
    \caption{Flux line cross-coupling matrix for the four qubits on the $\pi$-flux rhombus device. Note that the color for the diagonal elements is saturated on the colorbar. The matrix for the zero-flux rhombus is qualitatively similar.}
    \label{fig:xc}
    \end{minipage}
\end{figure}

We characterize and compensate for both DC-flux crosstalk and distortions in the fast-flux pulse shape, as described in the Methods.
The measured DC-flux cross-coupling matrix for the $\pi$-flux device is shown in Fig.\,\ref{fig:xc}; we note that the zero-flux device exhibits a similar matrix.

\begin{figure}
    \centering
    \includegraphics{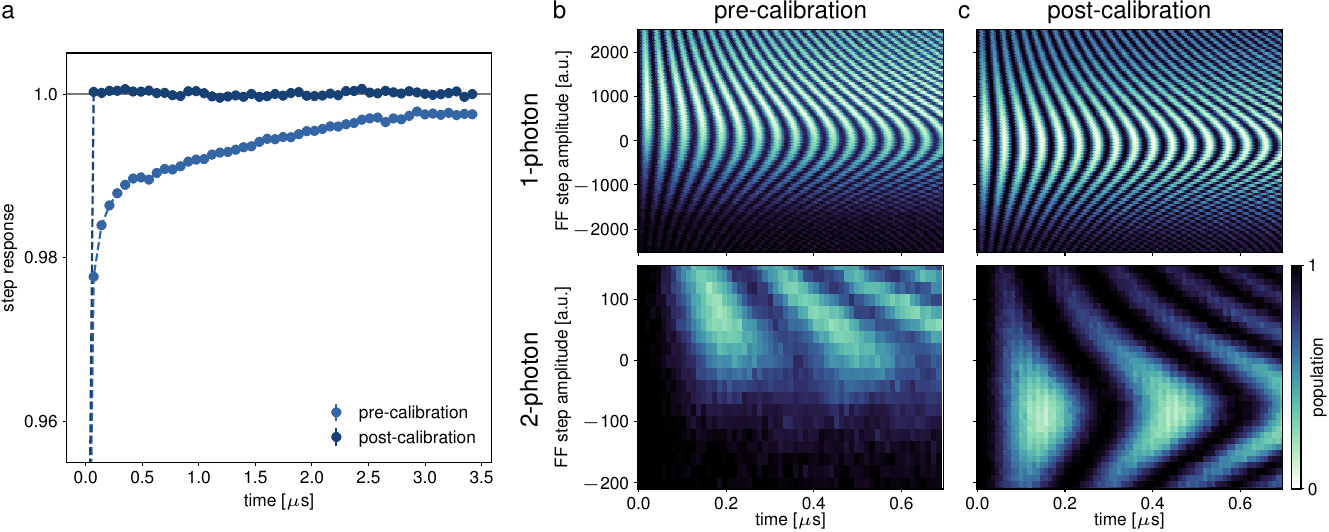}
    \caption{Fast-flux calibration for $\pi$-flux rhombus device, top qubit. \textbf{(a)} Spectroscopically measured qubit response as a function of time for pre- and post-calibration fast-flux pulses. Dashed lines are included as a guide to the eye and connect the experimentally measured points. The horizontal gray line is also included as a guide to the eye. \textbf{(b)} (upper) Swapping of qubit excitation between the top and left qubits, where the top qubit is initialized in the first excited state. After initialization, a step function is applied to the fast-flux line of the top qubit with varying amplitude (``FF step amplitude'', $y$-axis). The time-dependent fast-flux amplitude corresponding to the slowest swap over time, plus the reduced contrast of off-resonance swapping, indicates that the fast-flux pulse requires calibration. (lower) Same, but for a swap of the second excited state. \textbf{(c)} Same as (a), but with the calibrated fast-flux pulse.}
    \label{fig:FF}
\end{figure}

In Fig.\,\ref{fig:FF} we plot the top qubit response to the fast-flux pulse before and after its calibration.
We additionally evaluate the performance of fast-flux pulses through single-particle and doublon walks over two qubits, where the initially excited qubit is brought into resonance with the second qubit via the fast-flux pulse.
In these evaluations, the fast-flux pulse amplitude is varied and subsequent excitation population over time measured.
Because oscillations are the slowest when the qubits are on resonance, the single-qubit walk reveals imperfections in the fast-flux pulse when the optimal fast-flux amplitude appears to vary over time.
Similarly, asymmetric oscillation contrast for fast-flux setpoints above and below resonance reflects imperfections in the fast-flux pulse.
We find that both of these effects are qualitatively eliminated post-calibration.

\section{Readout Parameters}

\begin{figure}
    \centering
    \includegraphics{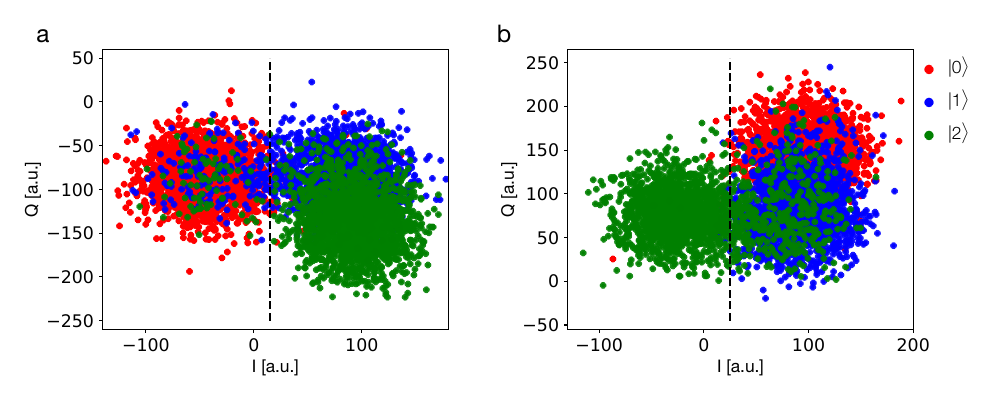}
    \caption{State-dependent quadrature I/Q measurements when optimizing for separability of different states. (a) A readout tone frequency of 7.2305 GHz optimizes differentiation of the $|0\rangle$ state from the $|1\rangle$ and $|2\rangle$ states. (b) A readout tone frequency of 7.2300 GHz isolates the $|2\rangle$ state best and is used for measuring doublon dynamics. In both cases, the averaging time for each measurement was 2.5 $\mu s$. Dashed line indicates the threshold used for assigning single-shot measurements to the qubit state populations.
    }
    \label{fig:SS}
\end{figure}

We determine qubit state occupancy using the state-dependent dispersive shift of the corresponding readout resonator.
To calibrate, we initialize the qubit into the $|0\rangle$, $|1\rangle$, or $|2\rangle$ state using $\pi$ pulses between adjacent levels and measure the resulting quadrature I/Q signal.
Readout tone frequencies, powers, and averaging times are separately optimized for discrimination between the $|0\rangle$ and $|1\rangle$ states, versus the $|1\rangle$ and $|2\rangle$ states.
In the single-particle and interacting particle-particle dynamics, we optimize for separability between the $|0\rangle$ and $|1\rangle$ state I/Q signals as seen in Fig.\,\ref{fig:SS}a.
For the doublon dynamics, we optimize for separability between the $|1\rangle$ and $|2\rangle$ states as seen in Fig.\,\ref{fig:SS}b. 

We use separately measured readout fidelities for each qubit to correct for readout error on that qubit.
Across all four qubits and both devices, the average readout fidelity for distinguishing the $|0\rangle$ from the $|1\rangle$ state is 86(2)\% while the average readout fidelity for distinguishing the $|2\rangle$ from the $|1\rangle$ state is 80(2)\%.
We note the reduced readout fidelity for the second excited state predominantly results from the lower $T_1$ coherence time for higher qubit levels.
Higher readout fidelities could be achieved using a near-quantum-limited amplifier \cite{macklin_nearquantum-limited_2015} and reduced measurement time. 

For simultaneous readout of two qubits, we perform a similar calibration where we initialize in the product states $|00\rangle$, $|01\rangle$, $|10\rangle$, and $|11\rangle$ and concurrently measure the resulting quadrature I/Q signals for the two corresponding readout resonators.
The two qubits are detuned from each other to suppress coupling during readout; in particular, nearest neighbors are detuned by at least 200\,MHz.
These readout frequencies differ from that of single-qubit readout; correspondingly, readout-tone frequencies and powers are reoptimized.
The measurement averaging time remains the same.
For all qubit pairs, each qubit's measurement fidelity is independent of the state of the other and consistent with its single-qubit measurement fidelity.

\section{Effective Dephasing Rates}

\begin{figure}
    \centering
    \includegraphics[height=\textwidth / 3]{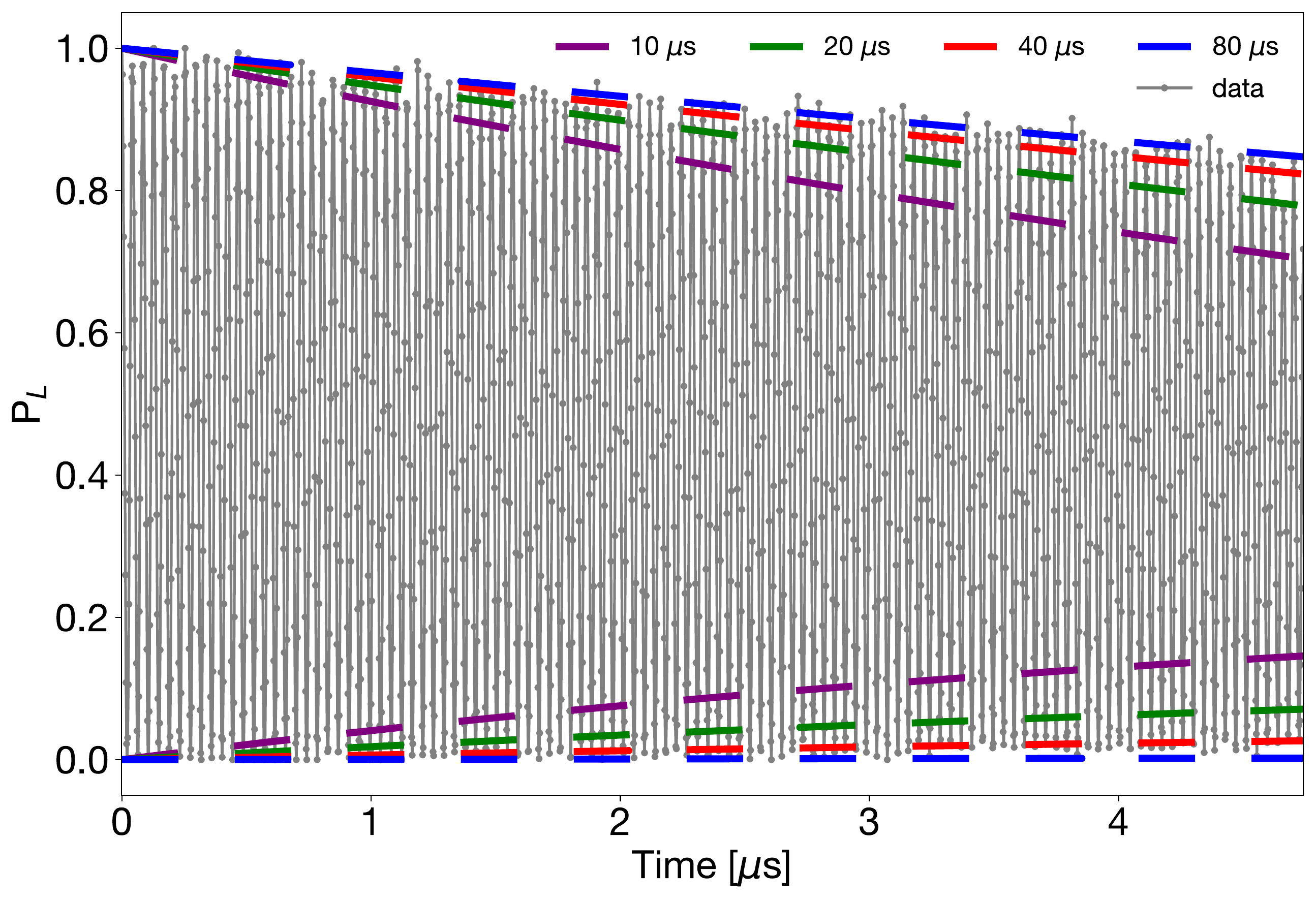}
    \caption{Single-photon oscillations between left and bottom qubits, with envelopes from the analytical model corresponding to various $T_\varphi$ dephasing times. Comparisons of the upper and lower bounds of the envelopes to data suggest 40\,$\mu s$ as a suitable lower bound for $T_\varphi$ and 80\,$\mu s$ as a reasonable upper bound.}
    \label{fig:T2eff}
\end{figure}

We estimate the effective $T_\varphi$ of our coupled system as in \cite{guo_observation_2021} because coupled systems have reduced flux dispersion compared to that of individual qubits.
To do this, we use single-photon oscillations between the zero-flux left (L) and bottom (B) qubits when both are tuned to the target qubit frequency. 
Under the simplification that both qubits have the same effective dephasing rate $T_{\varphi}$, the expected left-qubit first-excited-state population over time $\tau$ can be written as:
\begin{equation}
    P_L(\tau) = \frac{1}{2}\mathrm{cos}(2t\tau)\mathrm{exp}(-\frac{\tau}{2T_{1, L}} - \frac{\tau}{2T_{1, B}} - \frac{\tau}{T_{\varphi}}) + \frac{1}{2}\mathrm{exp}(-\frac{\tau}{2T_{1, L}} - \frac{\tau}{2T_{1, B}})
\end{equation}
for a coupling rate $t$ and depolarization times $T_{1,L}$ and $T_{1,B}$ for the left and bottom qubits, respectively.
With our independently measured $T_{1,L}$ and $T_{1,B}$ values, we vary the dephasing time and compare the exponentially decaying envelope of $P_L(\tau)$ to measurement as seen in Fig.\,\ref{fig:T2eff}.
We find that $T_\varphi$ values between 40\,$\mu s$ and 80\,$\mu s$ are consistent with the data.
Given that single-qubit $T_2$ times are on the higher end of the $T_2$ values across all eight qubits in both devices, we use 40\,$\mu s$ as the effective dephasing rate of the coupled system for both devices.
For doublon dynamics, we use $T_\varphi = 20 \mu s$ because the flux dispersion is approximately twice that of the ground-to-first-excited-state transition. 

\section{Interaction-induced coherent beating}

Here we elaborate on the observed beating of oscillations in the measured doublon populations of main text Fig.\,\ref{fig:2particle} and particle-particle populations of main text Fig.\,\ref{fig:interactingwalk}.

These are perhaps best understood through the energy eigenbases of these systems.
More specifically, the qualitative models presented in the main text treat the doublon subspace as a decoupled subspace from the particle-particle subspace; however, this is only true in the limit of infinite interaction strength.
For finite interactions, the energy eigenstates are hybridized between states in the two subspaces, albeit weakly for strong interactions.
Furthermore, this hybridization shifts the eigenstate energies.
Then, because the initial Fock states are superpositions of numerous energy eigenstates, time evolution under the Hamiltonian results in differing phase accumulation between the eigenstates, resulting in the coherent beating of population oscillations.
We note that the visibility of coherent beating is also a consequence of finite-size effects as there are simply fewer energy eigenstates to cause full dephasing.

\begin{figure}
    \centering
    \begin{minipage}{0.55\textwidth}
    \caption{Two-photon walk with single site detuned. \textbf{(a)} Experimental measurement of a two-photon walk on the $\pi$ plaquette taken over the same timescale as main text Fig.\,\ref{fig:2particle}. Notably, the amplitude of oscillations on sites L and R is larger at long times compared to the data presented in the main text. However, this behavior can be reproduced in simulation \textbf{(b)} by detuning the on-site energy of sites L or R by as little as $0.2$\,MHz. \textbf{(c)} Simulation of two-photon walk with all four qubits on resonance, identical to Fig.\,\ref{fig:2particle}b lower, for easy reference. }
    \label{fig:detunedsite}
    \end{minipage}
    \begin{minipage}{0.40\textwidth}
    \hfill
    \includegraphics[width=0.95\textwidth]{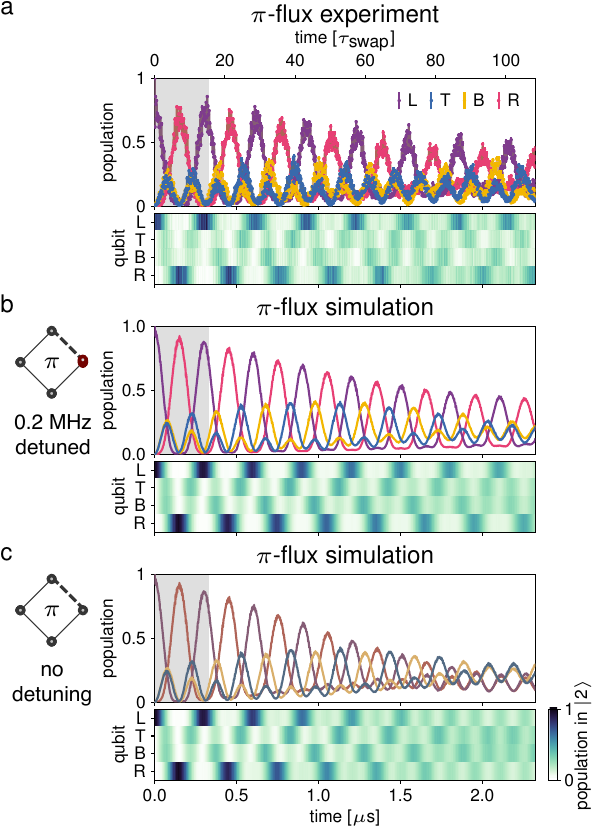}
    \end{minipage}
\end{figure}

We find in the bound two-photon walk, a weak detuning the left or right qubit significantly suppresses the $2\mu$s-timescale beating.
Fig.\,\ref{fig:detunedsite} shows the experimentally measured dynamics of a doublon with the right qubit detuned by $0.2$\,MHz, along with simulations with and without this detuning for comparison.
The sensitivity to detuning suggests that it breaks a degeneracy of the system or further perturbs the eigensystem in a way that reduces the observed beating effect.

\section{Photon-photon dynamics}

In the main text, we qualitatively represent the interacting particle-particle walk through the Hamiltonian adjacency graph.
Each state in the Fock-state subspace is treated as a site, with coupling strengths and signs inherited from the interacting two-particle Hamiltonian.
Time evolution of an initial Fock state in the Hamiltonian, then, corresponds to single-particle evolution on the graph, initialized on the corresponding site.
We note that such approaches can also be used to experimentally simulate interacting particle systems \cite{zhou_observation_2023}; however they scale unfavorably with particle number.

\begin{figure}
    \centering
    \begin{minipage}{0.5\textwidth}
    \includegraphics{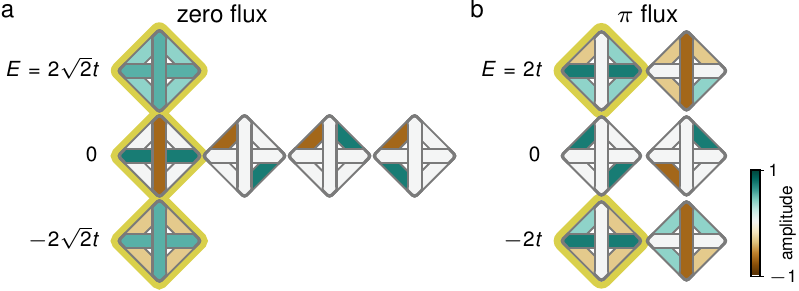}
    \end{minipage}
    \begin{minipage}{0.49\textwidth}
    \caption{Physics of photon-photon dynamics in the rhombus systems. \textbf{(a)} Two-particle eigenstates for the zero-flux plaquette, in the subspace where particles occupy different sites. Three eigenstates, outlined, exhibit nonzero probability amplitude of the $|\mathrm{LR}\rangle$ Fock state. \textbf{(b)} Same as (a), but for the $\pi$-flux plaquette. Here, only two eigenstates exhibit nonzero probability amplitude of the $|\mathrm{LR}\rangle$ Fock state. 
    }
    \label{fig:photphotdyn}
    \end{minipage}
\end{figure}

An alternative approach to understanding the interacting particle-particle walk examines the eigenstates of the Fock-state subspace as depicted in Fig.\,\ref{fig:photphotdyn}.
The zero-flux plaquette eigenspectrum exhibits three energies, each of which has an eigenstate (outlined in yellow) that has nonzero overlap with the $|\mathrm{LR}\rangle$ Fock state: that is, nonzero overlap with the initial state of a particle on the left site and a particle on the right site.
Time evolution of the superposition of these states results in nonzero probability amplitudes of all Fock states in this subspace at some point in time. 
With $\pi$ flux, the eigenspectrum still exhibits three distinct energy values, only two of which have eigenstates that overlap with $|\mathrm{LR}\rangle$.
Furthermore, they both have zero probability amplitude of the $|\mathrm{TB}\rangle$ state. As a result, we do not expect to simultaneously observe a particle on the top site and a particle on the bottom site at any point in time (in the limit of zero disorder and dephasing).

\section{Increasing particle number}
	
\begin{figure}
		\centering
		\includegraphics[width=\textwidth]{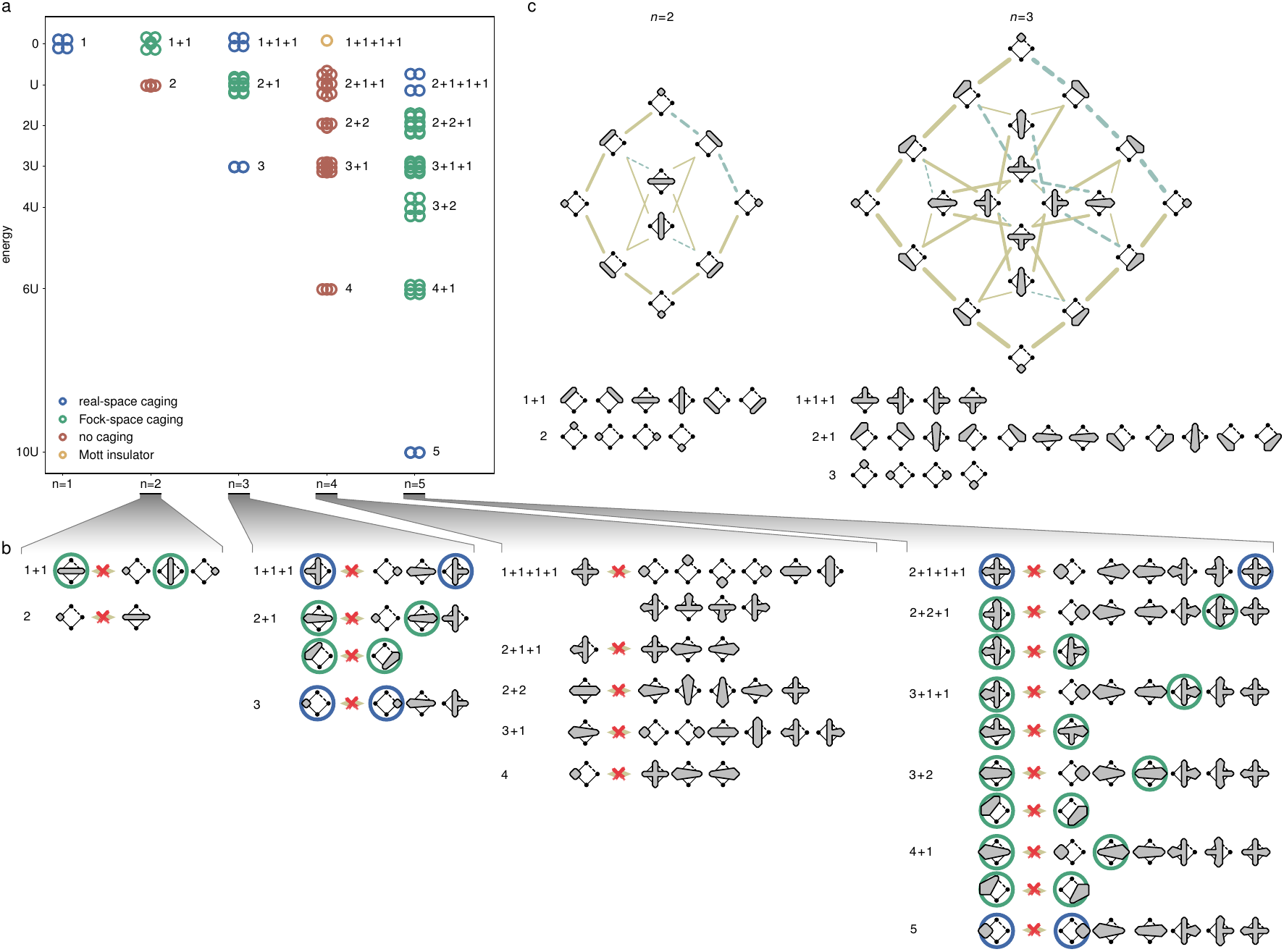}
		\caption{Adding particles to the $\pi$-rhombus plaquette.
			\textbf{(a)} Energy spectra for particle number $n=1$ to $n=5$. Points are offset horizontally slightly for visibility. Colors correspond to the presence or absence of caging within the corresponding approximate Fock-state partition subspace, in the hard-core limit (see text for details).
			\textbf{(b)} Tabulation of Fock states that are caged and the Fock states that they never reach, separated by blockaded arrows. 
			States that are outlined in blue exhibit real-space caging, while those outlined in green exhibit Fock-state caging. Caging from states that are neither highlighted nor outlined can be understood through the Hamitonian adjacency graph in (c). For simplicity, we do not include the caged states that can be mapped onto the ones shown under the symmetries of the plaquette.
			\textbf{(c)} Hamiltonian adjacency graphs for the entire Hilbert space for $n=2$ and $n=3$. Thicknesses of edges roughly indicate coupling strength, which differ due to bosonic enhancement;  yellow solid (blue dashed) color and linestyle indicate positive (negative) tunneling.
		}
		\label{fig:largerN}
	\end{figure}

	Here we consider particle number $n > 2$ on the $\pi$-rhombus plaquette.
	Figure \ref{fig:largerN}a shows the eigenenergies for $n=1$ to $n=5$ particles on the plaquette for a ratio of interaction strength to tunneling of $U/t = 13.5$, which is the approximate ratio of our experimental devices.
	For a given particle number, the eigenenergies are largely grouped around integer values of the interaction energy $U$.
	Correspondingly, the grouped eigenstates approximately reflect different Fock-state partitions of the $n$ particles over the four sites, labeled in the subfigure.
	For example, $n=3$ particles can be partitioned by placing all three particles on one site (``3'', with four Fock states); two particles on one site and one particle on a different site (``2+1'', with twelve Fock states), or all three particles on unique sites (``1+1+1'', with four Fock states).
	Because interactions are the dominant energy of the system, the respective eigenenergies are then close to $3U$, $U$, and $0$ (due to three, two, and no interacting particles on one of the sites).
	At the same time, we are not strictly in the hard-core limit of infinite interaction strength; as a result, these partition subspaces are not completely uncoupled and there is weak hybridization between them.
	In other words, the eigenstates labelled with a given partition do contain nonzero overlap with Fock states of different partitions.
	
	We numerically simulate the dynamics of initial Fock states within each of these partitions to examine whether real-space or Fock-space caging occurs.
	For finite interactions $U=13.5t$, we first consider dynamics across the entire Hilbert space.
	We find that within each Fock-space partition there exists at least one initial Fock state for which real-space or Fock-space caging occurs (see Fig.\,\ref{fig:largerN}b).
 This caging can be understood through examining the corresponding Hamiltonian adjacency graphs for the entire Hilbert space; Fig.\,\ref{fig:largerN}c shows the graphs for $n=2$ and $n=3$.
	Furthermore, this ubiquitous caging property persists for all nonzero interaction strengths.

	Next we consider the hard-core limit: for a given initial Fock state, we determine whether it explores all other Fock states \emph{within the partition subspace}.
	If at least one Fock state within a partition experiences caging, the entire subspace is labelled as exhibiting the caging property.
	In these characterizations, we label real-space caging as caging that is evident through average single-site measurements over all four sites.
	Fock-space caging is then caging that is only evident through average two- or multi-site measurements over all four sites.
	For example, the 1+1+1 subspace can be understood in the hard-core limit by considering the motion of the single hole, rather than the three particles.
	In this case, the single hole experiences caging, and (as in main text Fig.\,2) measurements of a single site at a time are sufficient to identify such caging.
	The results are color-coded in Fig.\,\ref{fig:largerN}a and the states that are real-space (Fock-space) caged are outlined in blue (green) in Fig.\,\ref{fig:largerN}b.
	Here, we find a general pattern where odd values of $n$ exhibit some form of caging within each subspace, while even values of $n$ largely do not, a possible generalization from the single-particle caging and doublon escape discussed in the main text.

	The real-space caged subspaces can be straightforwardly understood: to begin, the partitions of all particles on one site ($n$ odd) behave as a single composite particle that sees $\pi$ flux through the plaquette and are therefore caged.
	The other subspaces can be understood in relation to the Mott insulator case for $n=4$ where particles cannot tunnel because every site is occupied.
	With this in mind, the real-space caging within the 2+1+1+1 subspace can be described by the motion on a single particle on the Mott insulator background; similarly, the real-spacing caging of the 1+1+1 subspace can be thought of as the caging of a hole placed in the Mott insulator.
	
	The Fock-space caged subspaces also contain a rich structure.
	For odd $n$, each caged state seems to be caged from exactly one other state in its subspace.
	Furthermore, the two states are related by a 180-degree rotation, which results from the symmetry and flux structure of the Hamiltonian adjacency graph.
	Interestingly, there are Fock-space caged subspaces for even $n$ as well, for example the 1+1 subspace for $n=2$.
	We predict the corresponding subspaces are also Fock-space caged for even $n$ where $n \equiv 2$ (mod $4$), as they are equivalent to two particles on a Mott insulator background (which exists for $n$ where $n\equiv 0$ (mod $4$)).


\begin{thebibliography}{10}
\expandafter\ifx\csname url\endcsname\relax
  \def\url#1{\texttt{#1}}\fi
\expandafter\ifx\csname urlprefix\endcsname\relax\def\urlprefix{URL }\fi
\providecommand{\bibinfo}[2]{#2}
\providecommand{\eprint}[2][]{\url{#2}}

\bibitem{altman_quantum_2021}
\bibinfo{author}{Altman, E.} \emph{et~al.}
\newblock \bibinfo{title}{Quantum {Simulators}: {Architectures} and
  {Opportunities}}.
\newblock \emph{\bibinfo{journal}{PRX Quantum}} \textbf{\bibinfo{volume}{2}},
  \bibinfo{pages}{017003} (\bibinfo{year}{2021}).

\bibitem{carusotto_photonic_2020}
\bibinfo{author}{Carusotto, I.} \emph{et~al.}
\newblock \bibinfo{title}{Photonic materials in circuit quantum
  electrodynamics}.
\newblock \emph{\bibinfo{journal}{Nature Physics}}
  \textbf{\bibinfo{volume}{16}}, \bibinfo{pages}{268--279}
  (\bibinfo{year}{2020}).

\bibitem{leykam_artificial_2018}
\bibinfo{author}{Leykam, D.}, \bibinfo{author}{Andreanov, A.} \&
  \bibinfo{author}{Flach, S.}
\newblock \bibinfo{title}{Artificial flat band systems: from lattice models to
  experiments}.
\newblock \emph{\bibinfo{journal}{Advances in Physics: X}}
  \textbf{\bibinfo{volume}{3}}, \bibinfo{pages}{1473052}
  (\bibinfo{year}{2018}).

\bibitem{kollar_hyperbolic_2019}
\bibinfo{author}{Kollár, A.~J.}, \bibinfo{author}{Fitzpatrick, M.} \&
  \bibinfo{author}{Houck, A.~A.}
\newblock \bibinfo{title}{Hyperbolic lattices in circuit quantum
  electrodynamics}.
\newblock \emph{\bibinfo{journal}{Nature}} \textbf{\bibinfo{volume}{571}},
  \bibinfo{pages}{45--50} (\bibinfo{year}{2019}).

\bibitem{roushan_spectroscopic_2017}
\bibinfo{author}{Roushan, P.} \emph{et~al.}
\newblock \bibinfo{title}{Spectroscopic signatures of localization with
  interacting photons in superconducting qubits}.
\newblock \emph{\bibinfo{journal}{Science}} \textbf{\bibinfo{volume}{358}},
  \bibinfo{pages}{1175--1179} (\bibinfo{year}{2017}).

\bibitem{ma_dissipatively_2019}
\bibinfo{author}{Ma, R.} \emph{et~al.}
\newblock \bibinfo{title}{A dissipatively stabilized {Mott} insulator of
  photons}.
\newblock \emph{\bibinfo{journal}{Nature}} \textbf{\bibinfo{volume}{566}},
  \bibinfo{pages}{51--57} (\bibinfo{year}{2019}).

\bibitem{saxberg_disorder-assisted_2022}
\bibinfo{author}{Saxberg, B.} \emph{et~al.}
\newblock \bibinfo{title}{Disorder-assisted assembly of strongly correlated
  fluids of light}.
\newblock \emph{\bibinfo{journal}{Nature}} \textbf{\bibinfo{volume}{612}},
  \bibinfo{pages}{435--441} (\bibinfo{year}{2022}).

\bibitem{yan_strongly_2019}
\bibinfo{author}{Yan, Z.} \emph{et~al.}
\newblock \bibinfo{title}{Strongly correlated quantum walks with a 12-qubit
  superconducting processor}.
\newblock \emph{\bibinfo{journal}{Science}} \textbf{\bibinfo{volume}{364}},
  \bibinfo{pages}{753--756} (\bibinfo{year}{2019}).

\bibitem{gong_quantum_2021}
\bibinfo{author}{Gong, M.} \emph{et~al.}
\newblock \bibinfo{title}{Quantum walks on a programmable two-dimensional
  62-qubit superconducting processor}.
\newblock \emph{\bibinfo{journal}{Science}} \textbf{\bibinfo{volume}{372}},
  \bibinfo{pages}{948--952} (\bibinfo{year}{2021}).

\bibitem{karamlou_quantum_2022}
\bibinfo{author}{Karamlou, A.~H.} \emph{et~al.}
\newblock \bibinfo{title}{Quantum transport and localization in 1d and 2d
  tight-binding lattices}.
\newblock \emph{\bibinfo{journal}{npj Quantum Information}}
  \textbf{\bibinfo{volume}{8}}, \bibinfo{pages}{35} (\bibinfo{year}{2022}).

\bibitem{vidal_aharonov-bohm_1998}
\bibinfo{author}{Vidal, J.}, \bibinfo{author}{Mosseri, R.} \&
  \bibinfo{author}{Douçot, B.}
\newblock \bibinfo{title}{Aharonov-{Bohm} {Cages} in {Two}-{Dimensional}
  {Structures}}.
\newblock \emph{\bibinfo{journal}{Physical Review Letters}}
  \textbf{\bibinfo{volume}{81}}, \bibinfo{pages}{5888--5891}
  (\bibinfo{year}{1998}).

\bibitem{vidal_interaction_2000}
\bibinfo{author}{Vidal, J.}, \bibinfo{author}{Douçot, B.},
  \bibinfo{author}{Mosseri, R.} \& \bibinfo{author}{Butaud, P.}
\newblock \bibinfo{title}{Interaction {Induced} {Delocalization} for {Two}
  {Particles} in a {Periodic} {Potential}}.
\newblock \emph{\bibinfo{journal}{Physical Review Letters}}
  \textbf{\bibinfo{volume}{85}}, \bibinfo{pages}{3906--3909}
  (\bibinfo{year}{2000}).

\bibitem{lieb_two_1989}
\bibinfo{author}{Lieb, E.~H.}
\newblock \bibinfo{title}{Two theorems on the {Hubbard} model}.
\newblock \emph{\bibinfo{journal}{Physical Review Letters}}
  \textbf{\bibinfo{volume}{62}}, \bibinfo{pages}{1201--1204}
  (\bibinfo{year}{1989}).

\bibitem{tasaki_nagaokas_1998}
\bibinfo{author}{Tasaki, H.}
\newblock \bibinfo{title}{From {Nagaoka}'s {Ferromagnetism} to {Flat}-{Band}
  {Ferromagnetism} and {Beyond}: {An} {Introduction} to {Ferromagnetism} in the
  {Hubbard} {Model}}.
\newblock \emph{\bibinfo{journal}{Progress of Theoretical Physics}}
  \textbf{\bibinfo{volume}{99}}, \bibinfo{pages}{489--548}
  (\bibinfo{year}{1998}).

\bibitem{von_klitzing_quantum_2017}
\bibinfo{author}{von Klitzing, K.}
\newblock \bibinfo{title}{Quantum {Hall} {Effect}: {Discovery} and
  {Application}}.
\newblock \emph{\bibinfo{journal}{Annual Review of Condensed Matter Physics}}
  \textbf{\bibinfo{volume}{8}}, \bibinfo{pages}{13--30} (\bibinfo{year}{2017}).

\bibitem{bergholtz_topological_2013}
\bibinfo{author}{Bergholtz, E.~J.} \& \bibinfo{author}{Liu, Z.}
\newblock \bibinfo{title}{Topological flat band models and fractional {Chern}
  insulators}.
\newblock \emph{\bibinfo{journal}{International Journal of Modern Physics B}}
  \textbf{\bibinfo{volume}{27}}, \bibinfo{pages}{1330017}
  (\bibinfo{year}{2013}).

\bibitem{parameswaran_fractional_2013}
\bibinfo{author}{Parameswaran, S.~A.}, \bibinfo{author}{Roy, R.} \&
  \bibinfo{author}{Sondhi, S.~L.}
\newblock \bibinfo{title}{Fractional quantum {Hall} physics in topological flat
  bands}.
\newblock \emph{\bibinfo{journal}{Comptes Rendus Physique}}
  \textbf{\bibinfo{volume}{14}}, \bibinfo{pages}{816--839}
  (\bibinfo{year}{2013}).

\bibitem{spanton_observation_2018}
\bibinfo{author}{Spanton, E.~M.} \emph{et~al.}
\newblock \bibinfo{title}{Observation of fractional {Chern} insulators in a van
  der {Waals} heterostructure}.
\newblock \emph{\bibinfo{journal}{Science}} \textbf{\bibinfo{volume}{360}},
  \bibinfo{pages}{62--66} (\bibinfo{year}{2018}).

\bibitem{cao_unconventional_2018}
\bibinfo{author}{Cao, Y.} \emph{et~al.}
\newblock \bibinfo{title}{Unconventional superconductivity in magic-angle
  graphene superlattices}.
\newblock \emph{\bibinfo{journal}{Nature}} \textbf{\bibinfo{volume}{556}},
  \bibinfo{pages}{43--50} (\bibinfo{year}{2018}).

\bibitem{xie_fractional_2021}
\bibinfo{author}{Xie, Y.} \emph{et~al.}
\newblock \bibinfo{title}{Fractional {Chern} insulators in magic-angle twisted
  bilayer graphene}.
\newblock \emph{\bibinfo{journal}{Nature}} \textbf{\bibinfo{volume}{600}},
  \bibinfo{pages}{439--443} (\bibinfo{year}{2021}).

\bibitem{peotta_superfluidity_2015}
\bibinfo{author}{Peotta, S.} \& \bibinfo{author}{Törmä, P.}
\newblock \bibinfo{title}{Superfluidity in topologically nontrivial flat
  bands}.
\newblock \emph{\bibinfo{journal}{Nature Communications}}
  \textbf{\bibinfo{volume}{6}}, \bibinfo{pages}{8944} (\bibinfo{year}{2015}).

\bibitem{julku_quantum_2021}
\bibinfo{author}{Julku, A.}, \bibinfo{author}{Bruun, G.~M.} \&
  \bibinfo{author}{Törmä, P.}
\newblock \bibinfo{title}{Quantum {Geometry} and {Flat} {Band}
  {Bose}-{Einstein} {Condensation}}.
\newblock \emph{\bibinfo{journal}{Physical Review Letters}}
  \textbf{\bibinfo{volume}{127}}, \bibinfo{pages}{170404}
  (\bibinfo{year}{2021}).

\bibitem{goda_inverse_2006}
\bibinfo{author}{Goda, M.}, \bibinfo{author}{Nishino, S.} \&
  \bibinfo{author}{Matsuda, H.}
\newblock \bibinfo{title}{Inverse {Anderson} {Transition} {Caused} by
  {Flatbands}}.
\newblock \emph{\bibinfo{journal}{Physical Review Letters}}
  \textbf{\bibinfo{volume}{96}}, \bibinfo{pages}{126401}
  (\bibinfo{year}{2006}).

\bibitem{danieli_many-body_2020}
\bibinfo{author}{Danieli, C.}, \bibinfo{author}{Andreanov, A.} \&
  \bibinfo{author}{Flach, S.}
\newblock \bibinfo{title}{Many-{Body} {Flatband} {Localization}}.
\newblock \emph{\bibinfo{journal}{Physical Review B}}
  \textbf{\bibinfo{volume}{102}}, \bibinfo{pages}{041116}
  (\bibinfo{year}{2020}).

\bibitem{kuno_multiple_2021}
\bibinfo{author}{Kuno, Y.}, \bibinfo{author}{Mizoguchi, T.} \&
  \bibinfo{author}{Hatsugai, Y.}
\newblock \bibinfo{title}{Multiple quantum scar states and emergent slow
  thermalization in a flat-band system}.
\newblock \emph{\bibinfo{journal}{Physical Review B}}
  \textbf{\bibinfo{volume}{104}}, \bibinfo{pages}{085130}
  (\bibinfo{year}{2021}).

\bibitem{regnault_catalogue_2022}
\bibinfo{author}{Regnault, N.} \emph{et~al.}
\newblock \bibinfo{title}{Catalogue of flat-band stoichiometric materials}.
\newblock \emph{\bibinfo{journal}{Nature}} \textbf{\bibinfo{volume}{603}},
  \bibinfo{pages}{824--828} (\bibinfo{year}{2022}).

\bibitem{sutherland_localization_1986}
\bibinfo{author}{Sutherland, B.}
\newblock \bibinfo{title}{Localization of electronic wave functions due to
  local topology}.
\newblock \emph{\bibinfo{journal}{Physical Review B}}
  \textbf{\bibinfo{volume}{34}}, \bibinfo{pages}{5208--5211}
  (\bibinfo{year}{1986}).

\bibitem{aoki_hofstadter_1996}
\bibinfo{author}{Aoki, H.}, \bibinfo{author}{Ando, M.} \&
  \bibinfo{author}{Matsumura, H.}
\newblock \bibinfo{title}{Hofstadter butterflies for flat bands}.
\newblock \emph{\bibinfo{journal}{Physical Review B}}
  \textbf{\bibinfo{volume}{54}}, \bibinfo{pages}{R17296--R17299}
  (\bibinfo{year}{1996}).

\bibitem{yanay_two-dimensional_2020}
\bibinfo{author}{Yanay, Y.}, \bibinfo{author}{Braumüller, J.},
  \bibinfo{author}{Gustavsson, S.}, \bibinfo{author}{Oliver, W.~D.} \&
  \bibinfo{author}{Tahan, C.}
\newblock \bibinfo{title}{Two-dimensional hard-core {Bose}–{Hubbard} model
  with superconducting qubits}.
\newblock \emph{\bibinfo{journal}{npj Quantum Information}}
  \textbf{\bibinfo{volume}{6}}, \bibinfo{pages}{58} (\bibinfo{year}{2020}).

\bibitem{abilio_magnetic_1999}
\bibinfo{author}{Abilio, C.~C.} \emph{et~al.}
\newblock \bibinfo{title}{Magnetic {Field} {Induced} {Localization} in a
  {Two}-{Dimensional} {Superconducting} {Wire} {Network}}.
\newblock \emph{\bibinfo{journal}{Physical Review Letters}}
  \textbf{\bibinfo{volume}{83}}, \bibinfo{pages}{5102--5105}
  (\bibinfo{year}{1999}).

\bibitem{naud_aharonovbohm_2002}
\bibinfo{author}{Naud, C.} \emph{et~al.}
\newblock \bibinfo{title}{Aharonov–{Bohm} cages in the {GaAlAs}/{GaAs}
  system}.
\newblock \emph{\bibinfo{journal}{Physica E: Low-dimensional Systems and
  Nanostructures}} \textbf{\bibinfo{volume}{12}}, \bibinfo{pages}{190--196}
  (\bibinfo{year}{2002}).

\bibitem{mukherjee_experimental_2018}
\bibinfo{author}{Mukherjee, S.}, \bibinfo{author}{Di~Liberto, M.},
  \bibinfo{author}{Öhberg, P.}, \bibinfo{author}{Thomson, R.~R.} \&
  \bibinfo{author}{Goldman, N.}
\newblock \bibinfo{title}{Experimental {Observation} of {Aharonov}-{Bohm}
  {Cages} in {Photonic} {Lattices}}.
\newblock \emph{\bibinfo{journal}{Physical Review Letters}}
  \textbf{\bibinfo{volume}{121}}, \bibinfo{pages}{075502}
  (\bibinfo{year}{2018}).

\bibitem{hung_quantum_2021}
\bibinfo{author}{Hung, J.~S.} \emph{et~al.}
\newblock \bibinfo{title}{Quantum {Simulation} of the {Bosonic} {Creutz}
  {Ladder} with a {Parametric} {Cavity}}.
\newblock \emph{\bibinfo{journal}{Physical Review Letters}}
  \textbf{\bibinfo{volume}{127}}, \bibinfo{pages}{100503}
  (\bibinfo{year}{2021}).

\bibitem{li_aharonov-bohm_2022}
\bibinfo{author}{Li, H.} \emph{et~al.}
\newblock \bibinfo{title}{Aharonov-{Bohm} {Caging} and {Inverse} {Anderson}
  {Transition} in {Ultracold} {Atoms}}.
\newblock \emph{\bibinfo{journal}{Physical Review Letters}}
  \textbf{\bibinfo{volume}{129}}, \bibinfo{pages}{220403}
  (\bibinfo{year}{2022}).

\bibitem{karski_quantum_2009}
\bibinfo{author}{Karski, M.} \emph{et~al.}
\newblock \bibinfo{title}{Quantum {Walk} in {Position} {Space} with {Single}
  {Optically} {Trapped} {Atoms}}.
\newblock \emph{\bibinfo{journal}{Science}} \textbf{\bibinfo{volume}{325}},
  \bibinfo{pages}{174--177} (\bibinfo{year}{2009}).

\bibitem{preiss_strongly_2015}
\bibinfo{author}{Preiss, P.~M.} \emph{et~al.}
\newblock \bibinfo{title}{Strongly correlated quantum walks in optical
  lattices}.
\newblock \emph{\bibinfo{journal}{Science}} \textbf{\bibinfo{volume}{347}},
  \bibinfo{pages}{1229--1233} (\bibinfo{year}{2015}).

\bibitem{santos_methods_2020}
\bibinfo{author}{Santos, F. D.~R.} \& \bibinfo{author}{Dias, R.~G.}
\newblock \bibinfo{title}{Methods for the construction of interacting many-body
  {Hamiltonians} with compact localized states in geometrically frustrated
  clusters}.
\newblock \emph{\bibinfo{journal}{Scientific Reports}}
  \textbf{\bibinfo{volume}{10}}, \bibinfo{pages}{4532} (\bibinfo{year}{2020}).

\bibitem{kolovsky_conductance_2023}
\bibinfo{author}{Kolovsky, A.}, \bibinfo{author}{Muraev, P.} \&
  \bibinfo{author}{Flach, S.}
\newblock \bibinfo{title}{Conductance transition with interacting bosons in an
  {Aharonov}-{Bohm} cage}.
\newblock \emph{\bibinfo{journal}{arXiv}} \bibinfo{pages}{2303.00509}
  (\bibinfo{year}{2023}).

\bibitem{deng_superconducting_2016}
\bibinfo{author}{Deng, X.-H.}, \bibinfo{author}{Lai, C.-Y.} \&
  \bibinfo{author}{Chien, C.-C.}
\newblock \bibinfo{title}{Superconducting circuit simulator of {Bose}-{Hubbard}
  model with a flat band}.
\newblock \emph{\bibinfo{journal}{Physical Review B}}
  \textbf{\bibinfo{volume}{93}}, \bibinfo{pages}{054116}
  (\bibinfo{year}{2016}).

\bibitem{gneiting_lifetime_2018}
\bibinfo{author}{Gneiting, C.}, \bibinfo{author}{Li, Z.} \&
  \bibinfo{author}{Nori, F.}
\newblock \bibinfo{title}{Lifetime of flatband states}.
\newblock \emph{\bibinfo{journal}{Physical Review B}}
  \textbf{\bibinfo{volume}{98}}, \bibinfo{pages}{134203}
  (\bibinfo{year}{2018}).

\bibitem{chalker_anderson_2010}
\bibinfo{author}{Chalker, J.~T.}, \bibinfo{author}{Pickles, T.~S.} \&
  \bibinfo{author}{Shukla, P.}
\newblock \bibinfo{title}{Anderson localization in tight-binding models with
  flat bands}.
\newblock \emph{\bibinfo{journal}{Physical Review B}}
  \textbf{\bibinfo{volume}{82}}, \bibinfo{pages}{104209}
  (\bibinfo{year}{2010}).

\bibitem{longhi_inverse_2021}
\bibinfo{author}{Longhi, S.}
\newblock \bibinfo{title}{Inverse {Anderson} transition in photonic cages}.
\newblock \emph{\bibinfo{journal}{Optics Letters}}
  \textbf{\bibinfo{volume}{46}}, \bibinfo{pages}{2872-2875} (\bibinfo{year}{2021}).

\bibitem{biondi_incompressible_2015}
\bibinfo{author}{Biondi, M.}, \bibinfo{author}{van Nieuwenburg, E.~P.},
  \bibinfo{author}{Blatter, G.}, \bibinfo{author}{Huber, S.~D.} \&
  \bibinfo{author}{Schmidt, S.}
\newblock \bibinfo{title}{Incompressible {Polaritons} in a {Flat} {Band}}.
\newblock \emph{\bibinfo{journal}{Physical Review Letters}}
  \textbf{\bibinfo{volume}{115}}, \bibinfo{pages}{143601}
  (\bibinfo{year}{2015}).

\bibitem{katsura_mott_2021}
\bibinfo{author}{Katsura, H.}, \bibinfo{author}{Kawashima, N.},
  \bibinfo{author}{Morita, S.}, \bibinfo{author}{Tanaka, A.} \&
  \bibinfo{author}{Tasaki, H.}
\newblock \bibinfo{title}{Mott {Insulator}-like {Bose}-{Einstein}
  {Condensation} in a {Tight}-{Binding} {System} of {Interacting} {Bosons} with
  a {Flat} {Band}}.
\newblock \emph{\bibinfo{journal}{Physical Review Research}}
  \textbf{\bibinfo{volume}{3}}, \bibinfo{pages}{033190} (\bibinfo{year}{2021}).

\bibitem{flannigan_enhanced_2020}
\bibinfo{author}{Flannigan, S.} \& \bibinfo{author}{Daley, A.~J.}
\newblock \bibinfo{title}{Enhanced repulsively bound atom pairs in topological
  optical lattice ladders}.
\newblock \emph{\bibinfo{journal}{Quantum Science and Technology}}
  \textbf{\bibinfo{volume}{5}}, \bibinfo{pages}{045017} (\bibinfo{year}{2020}).

\bibitem{junemann_exploring_2017}
\bibinfo{author}{Jünemann, J.} \emph{et~al.}
\newblock \bibinfo{title}{Exploring {Interacting} {Topological} {Insulators}
  with {Ultracold} {Atoms}: {The} {Synthetic} {Creutz}-{Hubbard} {Model}}.
\newblock \emph{\bibinfo{journal}{Physical Review X}}
  \textbf{\bibinfo{volume}{7}}, \bibinfo{pages}{031057} (\bibinfo{year}{2017}).

\bibitem{kobayashi_superconductivity_2016}
\bibinfo{author}{Kobayashi, K.}, \bibinfo{author}{Okumura, M.},
  \bibinfo{author}{Yamada, S.}, \bibinfo{author}{Machida, M.} \&
  \bibinfo{author}{Aoki, H.}
\newblock \bibinfo{title}{Superconductivity in repulsively interacting fermions
  on a diamond chain: {Flat}-band-induced pairing}.
\newblock \emph{\bibinfo{journal}{Physical Review B}}
  \textbf{\bibinfo{volume}{94}}, \bibinfo{pages}{214501}
  (\bibinfo{year}{2016}).

\bibitem{mcclarty_disorder-free_2020}
\bibinfo{author}{McClarty, P.~A.}, \bibinfo{author}{Haque, M.},
  \bibinfo{author}{Sen, A.} \& \bibinfo{author}{Richter, J.}
\newblock \bibinfo{title}{Disorder-free localization and many-body quantum
  scars from magnetic frustration}.
\newblock \emph{\bibinfo{journal}{Physical Review B}}
  \textbf{\bibinfo{volume}{102}}, \bibinfo{pages}{224303}
  (\bibinfo{year}{2020}).

\bibitem{roy_interplay_2020}
\bibinfo{author}{Roy, N.}, \bibinfo{author}{Ramachandran, A.} \&
  \bibinfo{author}{Sharma, A.}
\newblock \bibinfo{title}{Interplay of disorder and interactions in a flat-band
  supporting diamond chain}.
\newblock \emph{\bibinfo{journal}{Physical Review Research}}
  \textbf{\bibinfo{volume}{2}}, \bibinfo{pages}{043395} (\bibinfo{year}{2020}).

\bibitem{salerno_interaction-induced_2020}
\bibinfo{author}{Salerno, G.}, \bibinfo{author}{Palumbo, G.},
  \bibinfo{author}{Goldman, N.} \& \bibinfo{author}{Di~Liberto, M.}
\newblock \bibinfo{title}{Interaction-induced lattices for bound states:
  {Designing} flat bands, quantized pumps, and higher-order topological
  insulators for doublons}.
\newblock \emph{\bibinfo{journal}{Physical Review Research}}
  \textbf{\bibinfo{volume}{2}}, \bibinfo{pages}{013348} (\bibinfo{year}{2020}).

\bibitem{tilleke_nearest_2020}
\bibinfo{author}{Tilleke, S.}, \bibinfo{author}{Daumann, M.} \&
  \bibinfo{author}{Dahm, T.}
\newblock \bibinfo{title}{Nearest {Neighbour} {Particle}-{Particle}
  {Interaction} in {Fermionic} {Quasi} {One}-{Dimensional} {Flat} {Band}
  {Lattices}}.
\newblock \emph{\bibinfo{journal}{Zeitschrift für Naturforschung A}}
  \textbf{\bibinfo{volume}{75}}, \bibinfo{pages}{393--402},(\bibinfo{year}{2020}).

\bibitem{khare_localized_2021}
\bibinfo{author}{Khare, R.} \& \bibinfo{author}{Choudhury, S.}
\newblock \bibinfo{title}{Localized dynamics following a quantum quench in a
  non-integrable system: an example on the sawtooth ladder}.
\newblock \emph{\bibinfo{journal}{Journal of Physics B: Atomic, Molecular and
  Optical Physics}} \textbf{\bibinfo{volume}{54}}, \bibinfo{pages}{015301}
  (\bibinfo{year}{2021}).

\bibitem{pyykkonen_flat_2021}
\bibinfo{author}{Pyykkönen, V. A.~J.} \emph{et~al.}
\newblock \bibinfo{title}{Flat band transport and {Josephson} effect through a
  finite-size sawtooth lattice}.
\newblock \emph{\bibinfo{journal}{Physical Review B}}
  \textbf{\bibinfo{volume}{103}}, \bibinfo{pages}{144519}
  (\bibinfo{year}{2021}).

\bibitem{stefanazzi_qick_2022}
\bibinfo{author}{Stefanazzi, L.} \emph{et~al.}
\newblock \bibinfo{title}{The {QICK} ({Quantum} {Instrumentation} {Control}
  {Kit}): {Readout} and control for qubits and detectors}.
\newblock \emph{\bibinfo{journal}{Review of Scientific Instruments}}
  \textbf{\bibinfo{volume}{93}}, \bibinfo{pages}{044709}
  (\bibinfo{year}{2022}).

\bibitem{braumuller_probing_2022}
\bibinfo{author}{Braumüller, J.} \emph{et~al.}
\newblock \bibinfo{title}{Probing quantum information propagation with
  out-of-time-ordered correlators}.
\newblock \emph{\bibinfo{journal}{Nature Physics}}
  \textbf{\bibinfo{volume}{18}}, \bibinfo{pages}{172--178}
  (\bibinfo{year}{2022}).

\bibitem{macklin_nearquantum-limited_2015}
\bibinfo{author}{Macklin, C.} \emph{et~al.}
\newblock \bibinfo{title}{A near–quantum-limited {Josephson} traveling-wave
  parametric amplifier}.
\newblock \emph{\bibinfo{journal}{Science}} \textbf{\bibinfo{volume}{350}},
  \bibinfo{pages}{307--310} (\bibinfo{year}{2015}).

\bibitem{guo_observation_2021}
\bibinfo{author}{Guo, Q.} \emph{et~al.}
\newblock \bibinfo{title}{Observation of energy-resolved many-body
  localization}.
\newblock \emph{\bibinfo{journal}{Nature Physics}}
  \textbf{\bibinfo{volume}{17}}, \bibinfo{pages}{234--239}
  (\bibinfo{year}{2021}).

\bibitem{johansson_qutip_2013}
\bibinfo{author}{Johansson, J.}, \bibinfo{author}{Nation, P.} \&
  \bibinfo{author}{Nori, F.}
\newblock \bibinfo{title}{{QuTiP} 2: {A} {Python} framework for the dynamics of
  open quantum systems}.
\newblock \emph{\bibinfo{journal}{Computer Physics Communications}}
  \textbf{\bibinfo{volume}{184}}, \bibinfo{pages}{1234--1240}
  (\bibinfo{year}{2013}).

\bibitem{zhou_observation_2023}
\bibinfo{author}{Zhou, X.}, \bibinfo{author}{Zhang, W.}, \bibinfo{author}{Sun,
  H.} \& \bibinfo{author}{Zhang, X.}
\newblock \bibinfo{title}{Observation of flat-band localization and topological
  edge states induced by effective strong interactions in electrical circuit
  networks}.
\newblock \emph{\bibinfo{journal}{Physical Review B}}
  \textbf{\bibinfo{volume}{107}}, \bibinfo{pages}{035152}
  (\bibinfo{year}{2023}).

\end{thebibliography}
\end{document}